\documentclass[reqno,11pt]{amsart}
\usepackage{graphicx}
\usepackage{slashed}
\usepackage{amscd}
\usepackage{tensor}
\usepackage{amssymb}
\usepackage{esint}
\usepackage{enumitem}
\usepackage{accents}
\usepackage[off]{auto-pst-pdf} 
\usepackage{pst-grad} 
\usepackage{pst-plot} 
\usepackage[mathscr]{eucal}
\numberwithin{equation}{section}
\allowdisplaybreaks[1]

\title[Dynamical Gravitational Coupling]{Dynamical Gravitational Coupling as a \\
Modified Theory of General Relativity}

\author[F.\ Finster]{Felix Finster}
\author[C.\ R\"oken]{Christian R\"oken \\ \\ April 2016}
\address{Fakult\"at f\"ur Mathematik \\ Universit\"at Regensburg \\ D-93040 Regensburg \\ Germany}
\email{finster@ur.de, Christian.Roeken@mathematik.ur.de}

\newtheorem{Def}{Definition}[section]
\newtheorem{Thm}[Def]{Theorem}
\newtheorem{Prp}[Def]{Proposition}
\newtheorem{Lemma}[Def]{Lemma}
\newtheorem{Remark}[Def]{Remark}

\newcommand{\Thanks}{\vspace*{.5em} \noindent \thanks}
\newcommand{\beq}{\begin{equation}}
\newcommand{\eeq}{\end{equation}}
\newcommand{\Proof}{\begin{proof}}
\newcommand{\QED}{\end{proof} \noindent}
\newcommand{\QEDrem}{\ \hfill $\Diamond$}

\newcommand{\la}{\langle}
\newcommand{\ra}{\rangle}

\newcommand{\R}{\mathbb{R}}
\newcommand{\1}{\mbox{\rm 1 \hspace{-1.05 em} 1}}

\renewcommand{\O}{{\mathscr{O}}}

\newcommand{\Dir}{{\mathcal{D}}}

\newcommand{\scrL}{\mycal L}
\newcommand{\scrM}{\mycal M}
\newcommand{\scrN}{\mycal N}
\newcommand{\scrS}{\mycal S}
\newcommand{\itemD}{\item[{\raisebox{0.125em}{\tiny $\blacktriangleright$}}]}

\newcommand{\vol}{\text{\rm{\tiny{vol}}}}
\DeclareFontFamily{OT1}{rsfso}{}
\DeclareFontShape{OT1}{rsfso}{m}{n}{ <-7> rsfso5 <7-10> rsfso7 <10-> rsfso10}{}
\DeclareMathAlphabet{\mycal}{OT1}{rsfso}{m}{n}
\DeclareMathOperator{\vleck}{\mathcal{V}}
\newcommand{\cliff}{\!\cdot\!}
\DeclareMathOperator{\grad}{grad}

\begin{document}

\maketitle

\begin{abstract}
A modified theory of general relativity is proposed, where the gravitational
constant is replaced by a dynamical variable in space-time.
The dynamics of the gravitational coupling is described by a family of parametrized null geodesics, 
implying that the gravitational coupling at a space-time point is determined by solving transport
equations along all null geodesics through this point.

General relativity with dynamical gravitational coupling (DGC) is introduced.
We motivate DGC from general considerations and explain how it
arises in the context of causal fermion systems.
The underlying physical idea is that the gravitational coupling is determined by
microscopic structures on the Planck scale which propagate with the speed of light.

In order to clarify the mathematical structure, we analyze the conformal behavior
and prove local existence and uniqueness of the time evolution.
The differences to Einstein's theory are worked out in the examples of
the Friedmann-Robertson-Walker model and the spherically symmetric collapse of a shell of matter.
Potential implications for the problem of dark matter and for inflation are discussed.
It is shown that the effects in the solar system are too small for being
observable in present-day experiments.
\end{abstract}

\tableofcontents

\section{Introduction}
Einstein's theory of general relativity is the basis of modern astrophysics and cosmology.
It was confirmed in various experiments (like the perihelion shift, the bending of light
and the red-shift effect for atomic clocks; see for example~\cite{Will:2014kxa}), and it led to
spectacular predictions like black holes, the big bang and gravitational waves
(see~\cite{misner, wald, straumann2}).
However, according to present cosmological models there are major discrepancies
between observations and Einstein's theory,
usually associated to {\em{dark matter}} and {\em{dark energy}} (see for
example~\cite{Trimble:1987ee, Peebles:2002gy}).
Instead of trying to fix this problem by introducing new and yet unobserved matter fields,
one can also think about resolving it by modifying the Einstein equations.
One idea, which goes back to Paul Dirac (see~\cite{dirac6} and \cite[Section~5]{dirac4})
is that the gravitational constant should be replaced by a dynamical variable in space-time.
This idea has been implemented in various modified theories of gravity,
notably in the scalar-tensor theories~\cite{Brans:1961sx, Brans:2005ra} 
and vector-tensor theories~\cite{Will:1972zz, Hellings:1973zz}.
In these theories, the dynamics of the gravitational coupling is described by
additional fields. The disadvantage of this procedure is that these additional
fields are introduced ad hoc, and that the description of the dynamics of these fields
involves additional free parameters.

Here we propose a different type of modification of the Einstein equations
in which the gravitational coupling becomes a dynamical variable.
However, in contrast to the above-mentioned scalar-tensor and vector-tensor theories,
our model of {\em{dynamical gravitational coupling}} (DGC) does not involve any free parameters.
In particular, DGC is not a ``deformation'' of general relativity, but it is a different theory.
It does not involve a parameter such that setting this parameter to zero
gives back Einstein's theory. Moreover, it is not formulated ad hoc,
but instead it follows naturally from concepts behind the theory of causal fermion systems,
being a recent proposal for a unified physical theory~\cite{cfs}.

Generally speaking, the DGC model is based on the
physical concept that on a microscopic length scale, space-time can no longer
be described by a Lorentzian manifold, but has a different non-trivial structure.
This microstructure should break the Lorentz invariance.
Considering macroscopic systems, however, the microstructure
should not be visible. Instead, it should be possible to describe the effective macroscopic behavior by
physical equations (like the Einstein equations, the Dirac equation or the Maxwell
equations) which are Lorentz invariant and are formulated geometrically on a Lorentzian manifold.

This physical picture is analogous to that of a crystal in solid state physics:
On the atomic scale, the crystal is composed of atoms, which are in a regular
configuration which distinguishes certain spatial directions. On scales much larger than
the atomic distance, however, the crystal can often be modelled by a homogeneous
and isotropic material, and it can be described effectively by macroscopic quantities like
temperature, pressure, densities, etc.

The effective macroscopic physical equations should be thought of as arising
by a suitable ``averaging process'' or ``homogenization process'' from the
physical equations which hold on the microscopic scale
(there is no consensus on how these microscopic physical equations should look like,
and at this stage we do not need to be specific).
In this averaging process, most information on the microscopic structure of space-time
gets lost. However, for what follows, it is important to take the point of view that
one parameter of the macroscopic theory should be a remnant of the microscopic theory:
the gravitational coupling constant. The simplest way to take this point of view is to identify
the Planck length with the length scale on which the microstructure becomes effective.
More generally, one can think of the Planck length as a quantity encoded in the
microstructure, but the length scale of the microstructure could for example be even much smaller
than the Planck length.

The considerations so far were very general and are widely accepted in the physics community.
A conclusion of these considerations is that, if the microscopic structure of space-time
has a dynamical behavior, then the gravitational constant should also be dynamical.
The crucial question is what the dynamics of the microstructure is and how the effective
behavior of the gravitational coupling can be described.
Here we make the assumption that the microstructure should have the same
dynamics as an ultraviolet regularization to a distributional solution of a hyperbolic
differential equation (like the wave or Dirac equation). Since the ultraviolet regularization modifies
the solution only on a microscopic scale, its dynamics is governed by the high-frequency behavior of the
hyperbolic equation and can thus be described by analyzing the propagation along
characteristics. In simple and more physical terms, we assume that the
microstructure propagates with the speed of light.
As will be explained in more detail in Section~\ref{secprep}, making this concept
precise leads quite naturally to DGC.

We remark that the above assumption that the microstructure should have the dynamics
of an ultraviolet regularization has a deeper justification in the theory of {\em{causal
fermion systems}}. In this theory, all space-time structures are encoded in an ensemble
of fermionic wave functions (the ``physical wave functions''). The physical equations
are formulated directly in terms of these wave functions via the so-called
{\em{causal action principle}}. In a certain limiting case (the ``continuum limit''), the Euler-Lagrange
equations of the causal action give rise to the Einstein equations (see~\cite[Section~4.9]{cfs}),
where the gravitational coupling is related to ultraviolet properties of the physical wave functions.
This will be explained in more detail in Section~\ref{seccfs}.

Since DGC should be of interest independent of the connection to
causal fermion systems, we here argue physically and derive the structure of DGC from general considerations.
This also has the benefit that the paper should be easily accessible to a broader
readership. Only the specific form of our DGC model as derived in Section~\ref{seccfs}
relies on results from the theory of causal fermion systems.
Consequently, since causal fermion systems are used only in Section~\ref{seccfs}, we 
decided not to give a general introduction to causal fermion systems and the causal
action principle, but refer the interested reader
to~\cite{cfs} or the survey papers~\cite{dice2014, intro}.

At present, the physical consequences and implications of DGC are difficult to estimate
because the analysis of physically realistic scenarios will take considerable time and effort.
Nevertheless, it is a main purpose of this paper to derive and discuss a few physical effects.
Our findings are summarized as follows: In a Friedmann-Robertson-Walker
space-time, the DGC function is proportional to the
square of the scale function (see Section~\ref{secfriedmann}).
Therefore, DGC suggests that in the early universe, the gravitational attraction
should have been weaker than measured at present. This bears some resemblance to
{\em{inflationary models}} which also reduce the strength of gravity in the early universe.
Moreover, our analysis of the spherically symmetric gravitational collapse (as studied in
Section~\ref{seccollapse} for a shell of matter) suggests that for a heavy collapsing
star, the gravitational coupling could in certain situations be larger than at present.
In such situations, the star would generate a stronger gravitational field than
in Einstein's theory. If this effect were sufficiently large, it could resolve the
problem of {\em{dark matter}}, because already the known matter would generate a sufficiently
strong gravitational field to give agreement with physical observations.
In the solar system, the effects of DGC seem too weak for being detectable in present-day experiments.

The paper is organized as follows. In Section~\ref{secmodel},
the DGC model is introduced, and it is explained
how to describe a constant gravitational coupling in Minkowski space.
Section~\ref{secderive} motivates the model from general considerations
and then gives a derivation in the context of causal fermion systems.
Section~\ref{secmath} is devoted to the mathematical structure of DGC.
Namely, the connection between conservation laws 
and conformal Killing symmetries as well as the behavior under conformal
transformations is worked out (Sections~\ref{seckilling} and~\ref{secconftrans}).
Moreover, we prove that the Einstein equations with DGC
coupled to usual matter fields (like a perfect fluid, Maxwell or Dirac field, etc.)
have a well-defined local time evolution (Section~\ref{seccauchy}).
In Section~\ref{secfriedmann}, the effect of DGC is analyzed in a
Friedmann-Robertson-Walker space-time.
Section~\ref{seccollapse} is devoted to the analysis of DGC in the gravitational
collapse of a star in the simplified model of a spherically symmetric shell of matter.
In Section~\ref{secweak}, linearized gravity and the Newtonian limit
are worked out, and potential effects in the solar system are discussed.
Finally, the appendices provide some additional background material:
Appendix~\ref{appcone} derives formulas for the volume and curvature of a
two-surface contained in the light cone. Appendix~\ref{apphadamard} introduces
the so-called regularized Hadamard expansion which puts the
physical arguments of Section~\ref{secderive} on a concise mathematical basis.
Finally, in Appendix~\ref{appE}, the form of the so-called DGC tensor is derived.

\section{General Relativity with Dynamical Gravitational Coupling} \label{secmodel}
For conceptual clarity, we now define general relativity with dynamical coupling.
The physical motivation, derivation and discussion will be given later in Section~\ref{secderive}.

\subsection{The Einstein Equations with Dynamical Gravitational Coupling} $ $\\
Let~$(\scrM, g)$ be a Lorentzian manifold. We also denote the Lorentzian metric
at a point~$p \in \scrM$ by~$\la u,v \ra_p = g_p(u,v)$ (where~$u,v \in T_p\scrM$).
A {\em{parametrized geodesic}} is a smooth mapping~$\gamma$ from an open interval~$I \subset \R$
to~$\scrM$ which satisfies the geodesic equation
\beq \label{nullgeodesic1}
\nabla_\tau \dot{\gamma}(\tau) = 0 \qquad \text{for all~$\tau \in I$}\:.
\eeq
A parametrized geodesic is {\em{null}} if it satisfies the relations
\beq \label{nullgeodesic2}
\big\la \dot{\gamma}(\tau), \dot{\gamma}(\tau) \big\ra_{\gamma(\tau)} = 0 \qquad \text{and} \qquad
\dot{\gamma}(\tau) \neq 0
\eeq
(it suffices to assume that these relations are satisfied for some~$\tau \in I$, because then the
geodesic equation implies that they hold for all~$\tau \in I$).
A null geodesic can be reparametrized in several ways. One obvious freedom is to
change the parameter by an {\em{additive constant}},
\beq \label{addchange}
\tilde{\gamma}(\tau) := \gamma(\tau+c) \qquad \text{with} \qquad
\tau \in \tilde{I} := I - c \:.
\eeq
Moreover, one can {\em{change the orientation}} of the geodesic, i.e.
\beq \label{fliporient}
\tilde{\gamma}(\tau) := \gamma(-\tau) \qquad \text{with} \qquad
\tau \in \tilde{I} := -I \:.
\eeq
If~$(\scrM, g)$ is time-orientable, one could distinguish an orientation and
restrict attention for example to future-directed null geodesics.
In order to keep our setting as general as possible, we prefer not to assume a
time orientation at this point.
Finally, one may reparametrize by a non-negative {\em{multiplicative constant}}~$\lambda$, i.e.
\beq \label{multchange}
\tilde{\gamma}(\tau) := \gamma(\lambda \tau) \qquad \text{with} \qquad
\text{$\tau \in \tilde{I} := I/\lambda$ and~$\lambda>0$} \:.
\eeq
This freedom scales the velocity vector by~$\dot{\tilde{\gamma}}(\tau) = \lambda\, \dot{\gamma}(\lambda \tau)$,
preserving both~\eqref{nullgeodesic1} and~\eqref{nullgeodesic2}.

In order to describe the dynamical behavior of the gravitational constant, we
want to distinguish the multiplicative reparametrization of all null geodesics.
Clearly, it suffices to consider maximal geodesics (i.e.\ geodesics which are inextendable),
because all other null geodesics can be obtained by restriction
to a smaller parameter domain.
\begin{Def} \label{defL}
We introduce a set of parametrized null geodesics
\beq \label{Gammadef}
\scrL = \big\{ (\gamma, I) \:|\: \text{$\gamma \::\: I \rightarrow \scrM$ is a
parametrized maximal null geodesic} \big\}
\eeq
with the following properties:
\begin{itemize}[leftmargin=2em]
\item[(a)] For every~$(\gamma, I) \in \scrL$, reparametrizing by an additive constant~\eqref{addchange}
or changing the orientation~\eqref{fliporient}  again gives a geodesic in~$\scrL$.
\item[(b)] For every maximal null geodesic~$\gamma(\tau)$ in~$\scrM$, there is exactly one~$\lambda>0$
such that the multiplicative reparametrization~\eqref{multchange} gives a
geodesic~$(\tilde{\gamma}, \tilde{I}) \in \scrL$.
\end{itemize}
\end{Def}

\begin{figure}
\begin{center}
\psscalebox{1.0 1.0} 
{
\begin{pspicture}(0,-2.2025)(13.535,2.2025)
\psellipse[linecolor=black, linewidth=0.02, dimen=outer](2.4525,-1.5125)(1.8125,0.105)
\psframe[linecolor=white, linewidth=0.02, fillstyle=solid, dimen=outer](5.235,-1.1975)(0.255,-1.5075)
\rput[bl](1.925,1.2125){\normalsize{$D_p\scrL$}}
\rput[bl](3.015,0.4125){\normalsize{$\big\{ u \in T_p\scrM \,|\, g_p(u,u)=0 \big\}$}}
\psline[linecolor=black, linewidth=0.02](0.65,2.0875)(4.25,-1.5125)
\psellipse[linecolor=black, linewidth=0.02, dimen=outer](2.4475,2.0975)(1.8125,0.105)
\psframe[linecolor=white, linewidth=0.02, dimen=outer](0.04,-2.1625)(0.0,-2.2025)
\psline[linecolor=black, linewidth=0.02](0.65,-1.5125)(4.25,2.0875)
\psbezier[linecolor=black, linewidth=0.04](1.165,1.5775)(1.2543558,1.5126325)(1.54607,1.0962533)(2.31,1.1475)(3.07393,1.1987468)(3.5639343,1.5088716)(3.86,1.7025)
\psbezier[linecolor=black, linewidth=0.04, linestyle=dashed, dash=0.17638889cm 0.10583334cm](1.18,1.5625)(1.1250777,1.7149678)(1.6626096,1.7753029)(1.865,1.7275)(2.0673904,1.6796972)(2.7650778,1.6749679)(3.01,1.7275)(3.2549224,1.7800322)(3.8823905,1.9396971)(3.87,1.7025)
\psbezier[linecolor=black, linewidth=0.04, linestyle=dashed, dash=0.17638889cm 0.10583334cm](3.745,-1.0175)(3.6556442,-0.9526324)(3.36393,-0.5362532)(2.6,-0.5875)(1.83607,-0.63874674)(1.3460658,-0.94887155)(1.05,-1.1425)
\psbezier[linecolor=black, linewidth=0.04](3.735,-0.9975)(3.7899222,-1.1499679)(3.2523904,-1.2103028)(3.05,-1.1625)(2.8476095,-1.1146972)(2.1499224,-1.1099678)(1.905,-1.1625)(1.6600777,-1.2150321)(1.0326096,-1.3746972)(1.045,-1.1375)
\rput[bl](2.15,-1.0925){\normalsize{$D_p\scrL$}}
\end{pspicture}
}
\end{center}
\caption{The two-surface~$D_p\scrL \subset T_p\scrM$.}
\label{figDpL}
\end{figure}
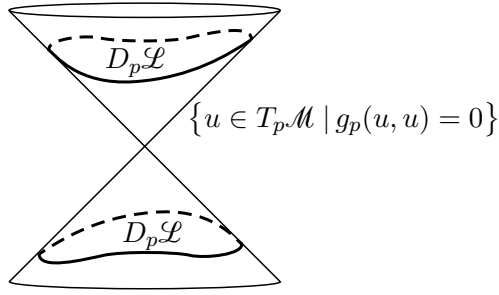

Next, for any space-time point~$p \in \scrM$ we introduce the set
\beq \label{DpLdef}
D_p\scrL = \big\{ \dot{\gamma}(\tau) \:\big|\: \text{$(\gamma, I) \in \scrL$, $\tau \in I$
and $\gamma(\tau) = p$} \big\} \subset T_p\scrM \:.
\eeq
Clearly, this set is a subset of the light cone centered at~$p$, i.e.
\[ D_p\scrL \subset \{ u \in T_p\scrM \,|\, \la u,u \ra_p = 0 \} \:. \]
Moreover, for every non-zero null vector~$u \in T_p\scrM$, the ray~$\R u$ intersects~$D_p\scrL$ in
exactly two points~$\pm \lambda u$ with~$\lambda>0$.
Assuming that~$\lambda$ depends smoothly on the null vector~$u$, the
set~$D_p\scrL$ is a smooth two-dimensional surface
consisting of two connected components, each of which is topologically a sphere
(see Figure~\ref{figDpL}).
The Lorentzian metric on~$T_p\scrM$ induces a Riemannian
metric on~$D_p\scrL$ (for details see Appendix~\ref{appcone}).
We denote the corresponding volume measure on~$D_p\scrL$ by~$d\mu_p$.
Furthermore, the Gaussian curvature (being one half scalar curvature) of~$D_p\scrL$ is denoted
by~$K_p \in C^\infty(D_p\scrL, \R)$.
We introduce the {\em{DGC function}}~$\kappa(p)$ at the space-time point~$p \in \scrM$ by
\beq \label{kappadyn}
\boxed{ \quad \frac{1}{\kappa(p)} = \frac{8 \pi}{\mu_p(D_p\scrL)} \int_{D_p\scrL} \frac{d\mu_p(x)}{K_p(x)} \:.
\quad }
\eeq
The {\em{Einstein equations with dynamical gravitational coupling}} and cosmological constant~$\Lambda$
are defined by
\beq \label{EDGC}
\boxed{ \quad \Big( R_{jk} - \frac{1}{2}\:R \: g_{jk} + \Lambda\, g_{jk} \Big)(p) = \kappa(p)\: T_{jk}(p)
+ E_{jk}(p) \:, \quad }
\eeq
where~$R_{jk}$ is the Ricci tensor of~$\scrM$, $R$ is scalar curvature, $\Lambda \in \R$
is the cosmological constant, and~$T_{jk}$ is the usual energy-momentum tensor of matter.
The tensor~$E_{jk}$, referred to as the {\em{DGC tensor}}, is a correction term needed in order to
get a mathematically consistent system of partial differential equations.
In order not to distract from the main ideas, the detailed form of~$E_{jk}$
will be stated and derived in Appendix~\ref{appE} (see~\eqref{Edef}). At this point, we prefer to make
a few remarks. First, the DGC tensor is trace-free, so that taking the trace of~\eqref{EDGC} gives
the equation
\beq \label{Etrace}
-R(p) + 4 \Lambda = \kappa(p)\: T(p) \:.
\eeq
Second, the DGC tensor is formed of
integrals of expressions involving~$T_{jk}$ and derivatives of~$\kappa$ along null geodesics.
In particular, if the energy-momentum tensor vanishes, then~$E_{jk}$ is also zero. Therefore,
in the vacuum, the above equations reduce to the usual vacuum Einstein equations~$R_{jk}=0$.
However, if matter is present, the space-time dependence of the DGC function~$\kappa(p)$ as
described by the dynamics of the null geodesics in~\eqref{kappadyn} comes into play and modifies
how matter generates curvature.

In this paper, we will restrict attention to the situation that~$\kappa(p)$ varies
only on a large scale. Then the DGC tensor should be very small,
implying that gravity with DGC should have all the properties of usual gravity,
except that the strength of the gravitational coupling depends on the space-time point~$p$.
With this in mind, our main objective is to understand how~$\kappa(p)$ depends on~$p$.

\subsection{A Simplified Model} \label{secsimp}
Since the form of the dynamical coupling function~\eqref{kappadyn} is rather involved
from the computational point of view,
we now propose a simpler model which is easier to analyze but still seems to capture 
the main features of~\eqref{kappadyn} and~\eqref{EDGC}. 

In order to derive the simplified model, let us assume that the Gaussian curvature is
a non-zero constant up to an error which only needs to be taken into account linearly, i.e.
\beq \label{Kex}
K_p(x) = C_p + e_p(x) \qquad \text{with} \qquad \big|e_p(x)\big| \ll C_p \:.
\eeq
Then, writing~\eqref{kappadyn} as
\[ \frac{1}{\kappa(p)} = 8 \pi\:\frac{\int_{D_p\scrL} K_p(x)^{-1}\: d\mu_p(x)}{\int_{D_p\scrL} \:d\mu_p(x)} \:, \]
we may insert any power of the Gaussian curvature into numerator and denominator without changing
the values of the integral, i.e.\ for any~$q \in \R$,
\beq \label{qform}
\frac{1}{\kappa(p)} = 8 \pi\: \frac{\int_{D_p\scrL} K_p^{q-1}(x) \:d\mu_p(x)}
{\int_{D_p\scrL} K_p^{q}(x)\:d\mu_p(x)} 
+ \O\big( e_p^2 \big) \:.
\eeq
Namely, substituting~\eqref{Kex} and expanding, we obtain
\begin{align*}
\frac{\int_{D_p\scrL} K_p^{q-1}(x) \:d\mu_p(x)}{
\int_{D_p\scrL} K_p^{q}(x)\:d\mu_p(x)} 
&= \frac{\int_{D_p\scrL} C_p^{q-2}\: \big(C_p + (q-1) \:e_p(x) \big) \:d\mu_p(x)}
{\int_{D_p\scrL} C_p^{q-1}\: \big(C_p + q \:e_p(x) \big) \:d\mu_p(x)} + \O\big( e_p^2 \big) \\
&= \frac{1}{C_p} - \frac{1}{C_p^2\, \mu_p(D_p\scrL)} \int_{D_p\scrL} e_p(x)\:d\mu_p(x)  + \O\big( e_p^2 \big) \:,
\end{align*}
which is obviously independent of~$q$.
Choosing~$q=1$, we may carry out the integral in the denominator in~\eqref{qform}
using the Gau{\ss}-Bonnet theorem.
Since~$D_p\scrL$ is topologically the disjoint union of two spheres, we obtain
\[ \int_{D_p\scrL} K_p(x)\:d\mu_p(x) = 8 \pi \:. \]
We conclude that
\[ \frac{1}{\kappa(p)} = \mu_p(D_p\scrL) + \O\big( e_p^2 \big) \:. \]
This expression is much simpler than~\eqref{kappadyn} because it becomes unnecessary
to compute the Gaussian curvature. Thus
for the simplified model we replace~$\kappa$ in~\eqref{EDGC} by
the DGC function~$\kappa_\vol$ defined by
\beq \label{kappavol}
\boxed{ \quad \frac{1}{\kappa_\vol(p)} = \mu_p(D_p\scrL) \quad }
\eeq
(clearly, we also replace~$\kappa(p)$ by~$\kappa_\vol(p)$ in the formula for the
DGC tensor~\eqref{Edef}).
Another advantage of this ansatz is that this expression is well-defined even
if~$D_p\scrL$ is not smooth; in fact it suffices that~$D_p \scrL$ is locally
an $L^2$-graph over the sphere (see Appendix~\ref{appcone}).

Using the H\"older inequality, one immediately sees that
under general assumptions, the DGC function
of the simplified model~\eqref{kappavol} is stronger than that
of the full model~\eqref{kappadyn}:
\begin{Lemma} \label{lemmaholder}
Assume that the Gaussian curvature~$K_p$ is everywhere non-negative on~$D_p \scrL$. Then
\[ \kappa(p) \leq \kappa_\vol(p) \:. \]
\end{Lemma}
\Proof Assume that~$K_p \geq 0$. Then the H\"older inequality implies that
\begin{align*}
\mu_p(D_p\scrL) &= \int_{D_p\scrL} \frac{\sqrt{K_p(x)}}{\sqrt{K_p(x)}}\: d\mu_p(x) \\
&\leq \bigg(\int_{D_p\scrL} K_p(x)\: d\mu_p(x) \bigg)^\frac{1}{2}
\bigg(\int_{D_p\scrL} \frac{1}{K_p(x)}\: d\mu_p(x) \bigg)^\frac{1}{2} \\
&= \sqrt{8 \pi} \;\bigg(\int_{D_p\scrL} \frac{1}{K_p(x)}\: d\mu_p(x) \bigg)^\frac{1}{2} \:.
\end{align*}
It follows that
\[ \int_{D_p\scrL} \frac{1}{K_p(x)}\: d\mu_p(x) \geq \frac{1}{8 \pi}\: \mu_p(D_p\scrL)^2 \:. \]
Applying~\eqref{kappadyn} and~\eqref{kappavol} gives the result.
\QED

\subsection{Example: Minkowski Space} \label{exmink}
Let us illustrate the above definitions in a simple example.
We let~$(\scrM, g)$ be Minkowski space. Then the null geodesics are straight lines of the form
\[ \gamma(\tau) = \tau u + x \qquad \text{with~$x,u \in \scrM$ and~$\la u,u \ra = 0$} \:. \]
Clearly, there are many ways to choose~$\scrL$. A simple method is to impose that~$u^0=\pm1/\varepsilon$
for a given parameter~$\varepsilon>0$. Thus we set
\beq \label{LMink}
\scrL = \Big\{ (\gamma, \R) \,\Big|\, \gamma(\tau) = \tau u + x \quad \text{with} \quad
\la u,u \ra = 0 \;\text{ and } \; \big|u^0 \big| = \frac{1}{\varepsilon} \Big\} .
\eeq
By direct computation, one finds that
\beq \label{dGMink}
D_p\scrL = \left\{ \pm \frac{1}{\varepsilon}\: \big( 1, \vec{n} \big) \quad \text{with} \quad
\vec{n} \in \R^3\:,\;
\big|\vec{n} \big| = 1  \right\} .
\eeq
This set consists of two $2$-spheres of radius~$1/\varepsilon$. Hence
the scalar curvature and the the volume of~$D_p\scrL$ are the constants
\beq \label{mupMink}
K_p = \varepsilon^2 \qquad \text{and} \qquad
\mu_p(D_p\scrL) = \frac{8 \pi}{\varepsilon^2} \:.
\eeq
Thus the dynamical coupling constant~\eqref{kappadyn} is also constant,
\beq \label{kappaMink}
\kappa(p) = \frac{\varepsilon^2}{8 \pi} \:.
\eeq
This shows that~$\varepsilon = 8 \pi \ell_P$ should be identified with a multiple of the Planck length.
The simplified model~\eqref{kappavol} also gives~$\kappa_\vol(p) = \varepsilon^2/(8 \pi)$,
giving agreement with~\eqref{kappaMink}.
As a consequence, the Einstein equations with DGC~\eqref{EDGC} reduce to the usual Einstein equations
with~$\kappa \sim \varepsilon^2$.
Hence the parameter~$\varepsilon$, which can be thought of as the Planck length, introduces a length scale
which determines the gravitational coupling constant.
We point out that the choice of~$\scrL$ clearly breaks Lorentz invariance
and distinguishes a specific reference frame.
Nevertheless, the Einstein equations~\eqref{EDGC} are Lorentz invariant (i.e.\ are tensor equations).
In other words, the violation of the Lorentz invariance by~$\scrL$ is not visible
in the equations of gravity.

By~\eqref{dGMink}, we realized the situation that the reference frame distinguished by
the regularization coincides with the rest frame of the considered physical system.
Later, we will also consider the situation that the rest frame of the physical system
is moving with constant velocity relative to the reference frame distinguished by the regularization.
In order to describe this situation, we let~$v$ be a future-directed timelike unit vector,
\beq \label{vdef}
\la v, v \ra = 1 \qquad \text{and} \qquad v^0>0 \:.
\eeq
We modify~\eqref{LMink} to
\beq \label{LMinkv}
\scrL = \Big\{ (\gamma, \R) \,\Big|\, \gamma(\tau) = \tau u + x \quad \text{with} \quad
\la u,u \ra = 0 \;\text{ and } \; \la u,v \ra = \pm \frac{1}{\varepsilon} \Big\} .
\eeq
Consequently, \eqref{dGMink} becomes
\beq \label{dGMinkv}
D_p\scrL = \Big\{ u \in \scrM \,\Big|\, \la u,u \ra=0 \text{ and } \la u,v \ra = \pm \frac{1}{\varepsilon} \bigg\} .
\eeq
Choosing~$v=(1,\vec{0})$, we clearly get back~\eqref{LMink} and~\eqref{dGMink}.
Moreover, by performing a Lorentz boost, it is obvious that the
right equation in~\eqref{mupMink} as well as~\eqref{kappaMink}
remain valid for general~$v$.

\section{Derivation of Dynamical Gravitational Coupling} \label{secderive}
We now explain how to come up with the Einstein equations with DGC.
To this end, we begin with general considerations (Section~\ref{secprep}) and then
give a derivation in the context of causal fermion systems (Section~\ref{seccfs}).

\subsection{General Considerations Leading to Dynamical Gravitational Coupling} \label{secprep}
We always work in natural units~$\hbar=c=1$. Then the masses of the elementary
particles give a length scale, the Compton scale (dividing by~$c=1$, we also have a corresponding time scale).
As long as gravity does not come into play,
the Compton scale determines the length scale in all physical processes.
Since time measurements also involve physical processes (for example in an atomic clock),
the Compton scale also enters the Lorentzian metric.
With this in mind, it is obvious and inevitable that the masses of the elementary particles are
constant in space-time (only mass ratios could change, but this will not be considered here).
Also, when we talk of length or time scales, these are always to be understood
relative to the Compton scale.

In the above units, the gravitational coupling has the dimension of length squared.
The resulting length scale is the {\em{Planck length}}~$\ell_P$,
\[ \ell_P = \sqrt{G} \approx 1.6 \times 10^{-35}\ \mathrm{m} \:. \]
It is generally believed that on length scales as tiny as the Planck length, the
conventional laws of physics should no longer hold, and yet unknown physical
effects should come into play.
Many physicists believe that on the Planck scale, the concept of the usual space-time
continuum breaks down, giving rise to yet unknown structures of a ``quantum space-time''
or ``quantum geometry.'' This idea is also commonly used in quantum field theory,
where the Planck length gives a natural length scale for the ultraviolet regularization,
thereby preventing the ultraviolet divergences of quantum field theory.
In what follows, we use these general concepts, but without specifying in detail
what the microscopic structure of space-time on the Planck scale should be.

A general idea behind DGC is that the microscopic space-time
structure should have a dynamical behavior, implying that the Planck length and
consequently also the gravitational constant should depend on the space-time point.
In order to describe the dynamics, we impose that the microstructure
{\em{propagates with the speed of light}}. This assumption can be motivated from the
above observation that the Planck length gives rise to an ultraviolet regularization and thus
affects the high-frequency behavior of the quantum fields. The high-frequency component of
a relativistic field, however, behaves typically like an ultrarelativistic wave and thus propagates
with the speed of light. A more detailed justification for working with light speed propagation
will be given in Section~\ref{seccfs} using concepts which arise in the context of
causal fermion systems.

\begin{figure}
\psscalebox{1.0 1.0} 
{
\begin{pspicture}(0,-3.1415234)(13.18,3.1415234)
\rput[bl](4.155,2.8834765){\normalsize{$t'$}}
\psframe[linecolor=white, linewidth=0.02, dimen=outer](3.82,-3.1015234)(3.78,-3.1415234)
\psbezier[linecolor=black, linewidth=0.06](4.03,0.35847655)(4.7821856,-0.2912303)(5.3680735,-0.8142134)(5.985,-1.0715234)(6.6019263,-1.3288335)(7.2167783,-1.3741452)(8.575,-2.5865235)
\psline[linecolor=black, linewidth=0.04, arrowsize=0.05291666666666668cm 2.0,arrowlength=1.4,arrowinset=0.0]{->}(8.43,-2.4515235)(8.43,-0.45152342)
\psline[linecolor=black, linewidth=0.04, arrowsize=0.05291666666666668cm 2.0,arrowlength=1.4,arrowinset=0.0]{->}(8.43,-2.4515235)(10.43,-2.4515235)
\psline[linecolor=black, linewidth=0.04, arrowsize=0.05291666666666668cm 2.0,arrowlength=1.4,arrowinset=0.0]{->}(4.03,0.34847656)(4.03,3.1484766)
\rput[bl](6.21,0.59347653){\normalsize{$x'$}}
\rput[bl](8.615,-0.5165234){\normalsize{$t$}}
\rput[bl](10.15,-2.8165236){\normalsize{$x$}}
\psbezier[linecolor=black, linewidth=0.02](3.675,1.1134765)(4.4837174,0.3919038)(5.145893,-0.4721343)(6.002714,-0.86896735)(6.8595347,-1.2658005)(7.364699,-1.2550405)(8.825,-2.6015234)
\psbezier[linecolor=black, linewidth=0.02](3.68,1.4734765)(4.4877005,0.76445657)(5.041663,-0.07363736)(5.886668,-0.59452206)(6.7316737,-1.1154068)(7.386535,-1.0884645)(8.845,-2.4115233)
\psbezier[linecolor=black, linewidth=0.02](3.9,1.6684766)(4.6521854,1.0187697)(5.3480735,-0.074213415)(6.16,-0.56652343)(6.9719267,-1.0588335)(7.476778,-0.9741452)(8.835,-2.1865234)
\psbezier[linecolor=black, linewidth=0.02](4.025,1.9384766)(4.7771854,1.257495)(5.4580736,0.21620834)(6.19,-0.33546755)(6.9219265,-0.88714343)(7.4867783,-0.7407853)(8.845,-2.0115235)
\psbezier[linecolor=black, linewidth=0.02](4.035,2.3334765)(5.187186,1.1987697)(5.5330734,0.5207866)(6.24,-0.03652344)(6.9469266,-0.59383345)(7.4717784,-0.56914514)(8.83,-1.7815235)
\psline[linecolor=black, linewidth=0.04](3.93,1.9534765)(4.13,1.9534765)
\psline[linecolor=black, linewidth=0.04](8.33,-0.85152346)(8.53,-0.85152346)
\psline[linecolor=black, linewidth=0.04](5.615,0.43347657)(5.615,0.23347656)
\psline[linecolor=black, linewidth=0.04](10.025,-2.3565235)(10.025,-2.5565233)
\psline[linecolor=black, linewidth=0.04, arrowsize=0.05291666666666668cm 2.0,arrowlength=1.4,arrowinset=0.0]{->}(4.035,0.34847656)(6.455,0.34847656)
\rput[bl](3.695,1.8484765){\normalsize{$1$}}
\rput[bl](8.6,-0.9665234){\normalsize{$1$}}
\rput[bl](9.94,-2.2115235){\normalsize{$1$}}
\rput[bl](5.06,-1.4715234){\normalsize{$\gamma(\tau)$}}
\rput[bl](0.0,0.58347654){\normalsize{$\grad_q \Gamma(p,q) = \dot{\gamma}(1)$}}
\psline[linecolor=black, linewidth=0.02, arrowsize=0.05291666666666668cm 2.0,arrowlength=1.4,arrowinset=0.0]{->}(4.045,0.35347655)(3.175,1.1334766)
\rput[bl](5.5,0.77347654){\normalsize{$1$}}
\psline[linecolor=black, linewidth=0.02, arrowsize=0.05291666666666668cm 2.0,arrowlength=1.4,arrowinset=0.0]{->}(5.74,-0.9665234)(4.375,-0.16652344)
\rput[bl](3.79,-0.67152345){\normalsize{$\dot{\gamma}(\tau)$}}
\pscircle[linecolor=black, linewidth=0.02, fillstyle=solid,fillcolor=black, dimen=outer](4.03,0.36347657){0.07}
\pscircle[linecolor=black, linewidth=0.02, fillstyle=solid,fillcolor=black, dimen=outer](5.72,-0.9515234){0.07}
\pscircle[linecolor=black, linewidth=0.02, fillstyle=solid,fillcolor=black, dimen=outer](8.435,-2.4515235){0.07}
\rput[bl](8.11,-2.6965234){\normalsize{$p$}}
\rput[bl](3.705,0.11347656){\normalsize{$q$}}
\pscircle[linecolor=black, linewidth=0.02, fillstyle=solid,fillcolor=black, dimen=outer](4.03,2.3434765){0.07}
\pscircle[linecolor=black, linewidth=0.02, fillstyle=solid,fillcolor=black, dimen=outer](8.435,-1.4365234){0.07}
\rput[bl](8.6,-1.4365234){\normalsize{$\tilde{p}(s)$}}
\rput[bl](4.185,2.3134766){\normalsize{$\tilde{q}(s)$}}
\rput[bl](5.1,1.2134765){\normalsize{$\tilde{\gamma}_s$}}
\end{pspicture}
}
\caption{Frequency shift as described by families of null geodesics.}
\label{fignull}
\end{figure}
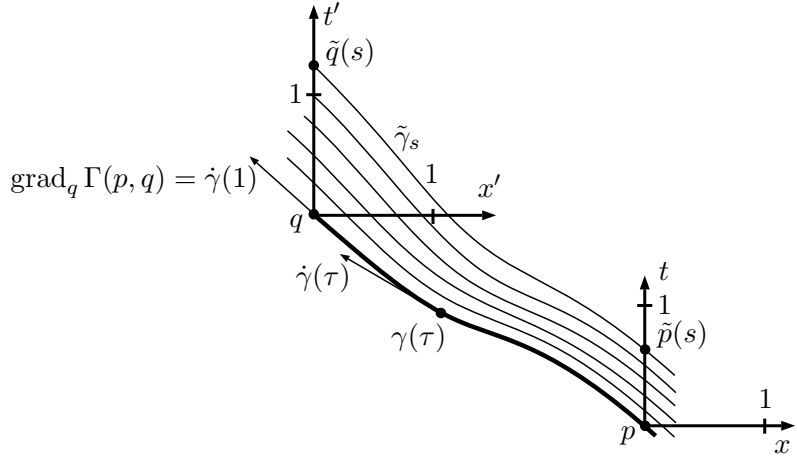

The behavior of length scales propagating with the speed of light can be
described most conveniently by families of null geodesics.
In order to explain the method in the simplest possible setting, we consider
a two-dimensional oriented and time-oriented space-time and let~$p$ and~$q$ be two space-time points
which can be joined by a null geodesic~$\gamma$ (see Figure~\ref{fignull}).
Assume that at the points~$p$ and~$q$, we have chosen local reference frames
(i.e.\ Gaussian coordinate systems) denoted by~$(t,x)$ and~$(t',x')$.
Next, we assume that at the point~$p$, we have a distinguished time scale~$\varepsilon_p$.
This time scale can be described by choosing a curve~$\tilde{p}(s)$, $s \in [0,1)$, moving
along the $t$-axis with velocity~$p'(s)$ given by
\beq \label{epspdef}
\tilde{p}(0) = p \qquad \text{and} \qquad \big\la \tilde{p}'(s), \tilde{p}'(s) \big\ra = \varepsilon_p^2\:.
\eeq
In order to obtain a corresponding scale at the point~$q$, we construct a family of null
geodesics~$(\tilde{\gamma}_s)_{s \in [0,1)}$ starting at~$\tilde{p}(s)$, i.e.
\[ \tilde{\gamma}_s(0) = \tilde{p}(s) \qquad \text{for all~$s \in [0,1)$} \:. \]
Assume that the null geodesic~$\tilde{\gamma}_s$ intersects the $t'$-axis at a point which we denote by~$\tilde{q}(s)$.
Then the resulting curve~$\tilde{q}(s)$ determines a time scale~$\varepsilon_q$ at~$q$,
as in analogy to~\eqref{epspdef} it satisfies the relations
\beq \label{epsqdef}
\tilde{q}(0) = q \qquad \text{and} \qquad \big\la \tilde{q}'(0), \tilde{q}'(0) \big\ra =: \varepsilon_q^2 \:.
\eeq
In this way, the dynamics of the time scale is described by analyzing the behavior of ``neighboring
null geodesics.'' In the example of Figure~\ref{fignull}, where neighboring null geodesics
are moving further apart, the resulting time scale~$\varepsilon_q$ at~$q$ will be larger
than the time scale~$\varepsilon_p$ at~$p$.

Before generalizing this concept to higher space-time dimensions, it is convenient to
formulate the connection between~$\varepsilon_p$ and~$\varepsilon_q$ in terms of the
geodesic distance function. To this end, we assume for simplicity that~$p$ and~$q$ are contained
in a geodesically convex subset~$\Omega$ of space-time,
meaning that any pair of points~$x,y \in \Omega$ can be joined by a unique geodesic in~$\Omega$
(see for example~\cite[Definition~1.3.2]{baer+ginoux}).
We let~$\Gamma(x,y)$ be the length of this geodesic squared, with the sign convention that~$\Gamma$ is positive
in timelike directions and negative in spacelike directions.
The fact that the points~$\tilde{p}(s)$ and~$\tilde{q}(s)$ both lie on the null geodesic~$\tilde{\gamma}_s$ implies that
\[ \Gamma \big( \tilde{p}(s), \tilde{q}(s) \big) = 0 \qquad \text{for all~$s \in [0,1)$}\:. \]
Differentiating at~$s=0$, we obtain
\[ 0 = \tilde{p}'(0)^j \:\frac{\partial}{\partial p^j} \Gamma(p,q) + \tilde{q}'(0)^j \:\frac{\partial}{\partial q^j} \Gamma(p,q) \:. \]
In order to simplify this formula, it is convenient to parametrize the null geodesic~$\gamma$ such that
\[ \gamma(0) = p \qquad \text{and} \qquad \gamma(1) = q \:. \]
Then the gradients of~$\Gamma(p,q)$ are multiples of the velocity vector of~$\gamma$, namely
\[ \grad_p \Gamma(p,q) = -2 \, \dot{\gamma}(0) \qquad \text{and} \qquad \grad_q \Gamma(p,q)
= 2 \, \dot{\gamma}(1) \]
(for the derivation see Lemma~\ref{lemmaGamma} in Appendix~\ref{apphadamard}).
We thus obtain
\beq \label{iprods}
\big\la \tilde{p}'(0), \dot{\gamma}(0) \big\ra = \big\la \tilde{q}'(0), \dot{\gamma}(1) \big\ra \:.
\eeq
Using that the vector~$\tilde{p}'(0)$ points in the time direction of the reference frame~$(t,x)$,
and similarly that~$\tilde{q}'(0)$ points in the time direction of the reference frame~$(t',x')$,
we can compute the inner products in~\eqref{iprods} with the help of~\eqref{epspdef} and~\eqref{epsqdef}.
This gives
\[ \varepsilon_p\: \dot{\gamma}(0)^0 = \varepsilon_q \: \dot{\gamma}(1)^0 \]
(where the index zero denotes the time component in the respective reference frame).
In a general affine parametrization of~$\gamma$, this identity can be written as
\[ \varepsilon_{\gamma(\tau)} \: \dot{\gamma}(\tau)^0 = \text{const} \]
(where the index zero now refers to the reference frame at~$\gamma(\tau)$).
For convenience, we parametrize the null geodesic such that the constant equals one, i.e.
\beq \label{epsdyn2d}
\varepsilon_{\gamma(\tau)}  = \frac{1}{\dot{\gamma}(\tau)^0} \qquad \text{for all~$\tau$}\:.
\eeq
We conclude that the regularization length changes like one over the time component
of the velocity vector of the null geodesic~$\gamma$.

In Figure~\ref{fignull}, we considered a null geodesic moving to the left. Clearly, one should also
consider null geodesics moving to the right. Therefore, one should not think of~$\varepsilon_p$ as
a scalar parameter, but as a function depending on the choice of the spatial direction.
Likewise, adapting the above consideration to a time-oriented four-dimensional space-time~$(\scrM, g)$,
the time scale~$\varepsilon_p$ depends on the choice of a spatial direction, and
its dynamics is described similar to~\eqref{epsdyn2d} in terms of the velocity vector of a null geodesic.
Taking into account all spatial directions systematically leads us to
introducing a set of parametrized null geodesics~$\scrL$ according to Definition~\ref{defL}.
Then the time scale~$\varepsilon_p$ in a reference frame~$(t, \vec{x})$ in a direction~$\vec{y} \neq 0$
at the space-time point~$p$ is obtained as follows: We choose a future-directed null geodesic
in~$\scrL$ with~$\gamma(0)=p$ such that the spatial component of~$\dot{\gamma}(0)$
is a positive multiple of~$\vec{y}$. Then, following~\eqref{epsdyn2d}, the time scale~$\varepsilon_p$
is given by
\beq \label{epsprel}
\varepsilon_p(\vec{y}) = \frac{1}{\dot{\gamma}(0)^0}\:.
\eeq

So far, we considered time-oriented space-times. In order to get rid of this assumption,
it is preferable to slightly modify the point of view.
Namely, instead of associating~$\varepsilon_p$ to a spatial direction~$\vec{y}$,
we shall work with a non-zero null vector~$u \in T_p\scrM$. Then, given such a null vector, we choose
a null geodesic in~$\scrL$ with~$\gamma(0)=p$ such that~$\dot{\gamma}(0)$
is a positive multiple of~$u$.
In the time-oriented case, this procedure is equivalent to that explained before~\eqref{epsprel}
because for every~$\vec{y}$ there is a unique time component~$t>0$ such that~$u = (t, \vec{y})$.
However, working with the null vector~$u$ has the advantage that it is no longer necessary to
distinguish a time direction. Indeed, the invariance of~$\scrL$ under changes of orientation~\eqref{fliporient}
ensures that~$\varepsilon_p$ remains unchanged if the sign of~$u$ is flipped.

Since the gravitational coupling constant~$\kappa$ is a scalar quantity of dimension length squared,
it must have the scaling~$\sim \varepsilon_p^2$.
Let us think how such a quantity can be obtained in our setting. Clearly, in order not to
distinguish a preferred reference frame, $\kappa$ should be formed geometrically out of the
two-surface~$\scrS := D_p \scrL$ in~\eqref{DpLdef}. On this surface, one has an induced Riemannian
metric~$d\mu_{\scrS}$ (for details see Appendix~\ref{appcone}).
Moreover, it is shown in Appendix~\ref{appcone} that the only curvature quantity on~$\scrS$
is the Gaussian curvature~$K_p$ (see Lemma~\ref{lemmaGauss}). Rescaling the velocity of the null geodesics
according to~$\dot{\gamma} \rightarrow \lambda \dot{\gamma}$, the volume measure and the
Gaussian curvature scales like~$d\mu_{\scrS} \rightarrow \lambda^2\, d\mu_{\scrS}$ and~$K_p \rightarrow K_p/\lambda^2$. In view of~\eqref{epsprel}, this means that the volume of~$\scrS$ has
the scaling dimension~$\varepsilon_p^2$, whereas the Gaussian curvature has the scaling
dimension~$\varepsilon_p^{-2}$.
This shows that in particular that our formulas for the DGC functions~\eqref{kappadyn} 
and~\eqref{kappavol} indeed have the correct scaling dimension
(i.e.\ the correct dimension of length squared).
But there are many other expressions with the correct scaling dimension.
For example, one can form powers of integrals of powers of scalar curvature
\[ \left( \int_\scrS K_p(x)^q\, d\mu_\scrS(x) \right)^\frac{2}{2-2q} \qquad \text{with~$q \in \R$}\:. \]
Using covariant derivatives of Gaussian curvature, one can also form expressions like
\[ \left( \int_\scrS \big|\Delta K_p(x) \big|^q\, d\mu_\scrS(x) \right)^\frac{2}{2-4q}, \quad 
\left( \int_\scrS \big|\nabla K_p(x) \big|^q\, d\mu_\scrS(x) \right)^\frac{2}{2-3q}, \quad \ldots \:. \]
The most general expression with the correct length dimension
is obtained by combining the above expressions
by iteratively applying the operation~$(x, y) \rightarrow x^\alpha\, y^{1-\alpha}$ with~$\alpha \in \R$.
In order to determine the form of dynamical coupling beyond such scaling arguments, we
will now use the detailed structure of the causal action principle used in the
theory of causal fermion systems.

\subsection{Derivation in the Context of Causal Fermion Systems} \label{seccfs}
The theory of {\em{causal fermion systems}} (see~\cite{cfs} and the references therein)
provides concise mathematical structures which should describe physical space-time on all length
scales. On the macroscopic scale,
these space-time structures reduce to the structures of a Lorentzian manifold.
However, on the Planck scale, causal fermion systems allow for non-trivial microstructures
which may be discrete or continuous.
The physical equations on a causal fermion systems are formulated via a novel
variational principle, referred to as the {\em{causal action principle}}.
For what follows, we only need a few key features of causal fermion systems, which we now
introduce. One feature is that all space-time structures (causal relations, geometric
structures, bosonic potentials, etc.) are encoded in an ensemble of quantum mechanical
wave functions defined the space-time points.
On the macroscopic scale, these wave functions are solutions of the Dirac equation.
The non-trivial microstructure is described by modifying the wave functions on the
microscopic scale, such that the Dirac equation no longer needs to hold.
In more technical terms, this is achieved by introducing an ultraviolet (UV) regularization
into the so-called kernel of the fermionic projector~$P(x,y)$.
Denoting the length scale of the regularization by a parameter~$\varepsilon>0$,
the dynamical behavior of the resulting regularized kernel~$P^\varepsilon(x,y)$
is described by the so-called {\em{regularized Hadamard expansion}}
as worked out in Appendix~\ref{apphadamard}.
This dynamics is consistent with the general considerations in the previous sections
and makes these considerations precise in the context of the Dirac equation.

Another feature of causal fermion systems of importance here is that
the Euler-Lagrange equations of the causal action give rise to
the Einstein equations in the so-called {\em{continuum limit}},
an approximation based on the assumption that the length scale~$\varepsilon$
of the regularization should be much smaller than the length scale~$\ell_{\text{macro}}$
of macroscopic physics (where by ``macroscopic physics'' we mean the length scales
accessible to experiments in high-energy physics).
The gravitational coupling is described by a length scale~$\delta$, which must be in the range
\[  \varepsilon \lesssim \delta  \ll \ell_\text{macro} \:. \]
We next make the following crucial assumption:
\beq \label{assumption}
\text{The dynamics on the scale~$\delta$ is the same as
that on the scale~$\varepsilon$.}
\eeq
This assumption is of course trivially satisfied if the length scales~$\delta$ and~$\varepsilon$ coincide.
However, if~$\delta$ defines another length scale which is much larger than~$\varepsilon$,
this assumption is questionable, as will be discussed in Remark~\ref{remcritical} below.
Making this assumption, the EL equations in the continuum limit give the equations
(see~\cite[Section~4.9]{cfs})
\beq \label{einstein}
\frac{C}{\varepsilon_p^2} \Big( R_{jk} - \frac{1}{2}\:R \: g_{jk} + \Lambda\, g_{jk} \Big)(p) = T_{jk}(p) \:,
\eeq
where~$C$ is a dimensionless constant (which in the case~$\delta \ll \varepsilon$ is much smaller than one),
and~$\varepsilon_p$ is the time scale in~\eqref{epsprel}. Since this time scale depends on the null
direction, the equation~\eqref{einstein} is to be evaluated weakly, meaning that we must take
``averages'' over the null directions. The natural way of doing so is to integrate both sides over~$D_p \scrL$.
Moreover, rewriting the factor~$1/\varepsilon_p^2$ geometrically gives a constant times $K_p^{-1}$.
We thus obtain the equations
\beq \label{einsteineff}
C' \:\bigg( \int_{D_p \scrL} \frac{d\mu_p(x)}{K_p(x)}\:  \bigg) 
\Big( R_{jk} - \frac{1}{2}\:R \: g_{jk} + \Lambda\, g_{jk} \Big)(p) = \mu_p \big( D_p \scrL \big)\:
T_{jk}(p)
\eeq
with a new constant~$C'$.
This procedure of bringing the equations into a geometrical form suffers from the shortcoming
that the resulting equations are no longer consistent (this is obvious because dividing~\eqref{einsteineff}
by~$\mu_p \big( D_p \scrL \big)$ and taking the divergence, the left side of the equation is non-zero,
but the right side vanishes). One method for obtaining consistent equations is to bring the equations
in variational form. Thus one would have to find a Lagrangian which is diffeomorphism invariant
such that the dominant terms of the resulting Euler-Lagrange equations coincide with~\eqref{einsteineff}.
This procedure would give a systematic method of deriving all the correction terms needed to
obtain consistent equations.

This variational approach also seems most convincing conceptually in view of the fact that
the equations~\eqref{einsteineff} are derived in~\cite{cfs} from a variational principle
(the causal actinon principle), suggesting
that the effective equations should again be of variational form
(for more details see~\cite[Section~4.7]{cfs}). Also, a variational form would be
desirable in order to make Noether's theorem applicable
(the connection between symmetries and conservation laws was established for
the causal action principle in~\cite{noether}).
However, bringing the Einstein equations with DGC in variational form seems to make
it necessary to also consider the family of geodesics~$\scrL$ as variable quantities
by formulating an action for the set~$\scrL$. This goes beyond the scope of the present paper.
For this reason, we simply add correction terms to~\eqref{einsteineff} which vanish if~$\kappa$
is constant and ensure that the resulting system of partial differential equations is 
mathematically consistent (for details see Appendix~\ref{appE}).

Clearly, the above derivation involved a few assumptions, which we now
discuss in detail.

\begin{Remark} {\bf{(critical discussion of assumption~\eqref{assumption})}} \label{remcritical} {\em{
The above derivation depends crucially on the assumption~\eqref{assumption}.
Therefore, we now discuss this assumption in detail.
The consideration of Figure~\ref{fignull} (and its higher-dimensional analog)
shows that the propagation along null geodesics describes the dynamics of a length scale.
Since this consideration is purely geometric, this dynamics is universal and applies
to all quantities which move along null geodesics.
However, the consideration no longer applies to quantities which propagate slower than the
speed of light. For example, mass parameters are constant in space-time, which is no
contradiction to our consideration because the mass changes the propagation speed.
The parameter~$\delta$ arises in~\cite{cfs} in order to describe two different
regularization effects: general surface states and shear states.
While the analogy between the curvature of the hypersurface describing the general surface states
and the curvature of the mass hyperbola suggests that~$\delta^{-1}$ should be regarded
as a very large mass parameter, it is not clear whether it is really constant in space-time.
On the contrary, it seems natural to assume that~$\delta$ is of the same order of magnitude as~$\varepsilon$,
in which case they should also have the same dynamical behavior.
For the shear states, there is no similarity to a mass, again suggesting a massless dynamics
and propagation with the speed of light.

To summarize, the theory of causal fermion systems strongly suggests DGC.
However, we would like to point out that DGC is not a compelling consequence
of the causal action principle. Assuming a massless dynamics for all regularization effects is an
interpolation which could only be justified by a more detailed knowledge or a better understanding
of the microscopic structure of space-time.
}} \QEDrem
\end{Remark}

\begin{Remark} {\bf{(alternative forms of the Einstein equations with DGC)}} \label{remalternative} {\em{
Rewriting the EL equations of the causal action in the continuum limit~\eqref{einsteineff}
involves some arbitrariness, as we now discuss:
\begin{itemize}[leftmargin=2em]
\item[(a)] In order to keep the setting as simple as possible,
we introduced~$\scrL$ as a set of null geodesics.
This description might be too simple, as we now explain.
In the context of causal fermion systems, the dynamics of the microstructure arises
by considering the high-frequency behavior of solutions of the Dirac equation.
Considering Dirac particles with an electric charge, this high-frequency limit is described by the
motion in the presence of the electromagnetic field,
\beq \label{EMpresence}
\nabla_\tau \dot{\gamma}^j = e F^j_{\;\;k}\: \dot{\gamma}^k \:.
\eeq
This suggests that, if electromagnetic fields are present, the geodesic equation~\eqref{nullgeodesic1}
should be replaced by~\eqref{EMpresence}.
This might have a relevant effect for example for a star in a strong magnetic field.
Similarly, the geodesic equation should also be modified if other fields (like strong or weak gauge fields)
are present.

The question if and how precisely the geodesic equation should be modified
depends crucially on the question which elementary particles
determine the gravitational coupling function. In~\cite[Chapter~4]{cfs} the gravitational constant is encoded
in the neutrino sector, giving an indication that the high-frequency limit of the neutrino
equation should be considered.
Consequently, only the weak gauge fields should be taken into account.
\item[(b)] Another potential modification is that also the cosmological constant~$\Lambda$ might have
a dynamical behavior. In the derivation in~\cite{cfs}, the parameter~$\Lambda$ in~\eqref{einsteineff}
can be chosen arbitrarily at every space-time point. In the Einstein equations~$R_{jk}-\frac{1}{2}\,R\, g_{jk}
+ \Lambda\, g_{jk} = 8 \pi G\, T_{jk}$ without DGC, taking the divergence
gives the equation~$\partial_j \Lambda=0$, implying that~$\Lambda$ must be constant in space-time.
But this argument no longer applies with DGC.
Therefore, it is conceivable that one should 
replace~$\Lambda$ in~\eqref{einsteineff} by a cosmological function~$\Lambda(x)$
(clearly, this would also make it necessary to modify the DGC tensor).

We remark that in principle, the dynamics of~$\Lambda$ could be derived from the causal action
by analyzing the corresponding EL equations to degree three on the light cone.
But this analysis would involve extensive computations which have not yet been carried out.
\item[(c)] The critical reader may wonder why we evaluated~\eqref{einstein} weakly
by integrating over the null directions. Should~\eqref{einstein} not be satisfied for all
null directions? 
We took the point of view that~\eqref{einstein} should be satisfied in the
strongest possible sense. Therefore, if a pointwise evaluation in every null direction
is impossible, we must take suitable ``averages'' over the null directions.
However, it is conceivable that the EL equations~\eqref{einstein} can indeed be satisfied
pointwise in every null direction if additional perturbations of the fermionic projector are taken into
account (this question could only be answered by a detailed and extensive analysis which
has not yet been carried out). If these additional perturbations were not dynamical
(similar to the microlocal chiral transformation in~\cite{cfs}), then this would have no effect
on our results, except that possibly one might have to modify
the ``averaging procedure'' in~\eqref{einsteineff}.
However, if these additional perturbations were dynamical (i.e.\ if they came with additional
hyperbolic field equations), this could change the picture completely.
In this case, one could hope that gravity could be described by a modification
or extension of the Einstein equations with DGC~\eqref{EDGC}.
\end{itemize}
Despite these potential extensions and modifications, it is fair to say that the
Einstein equations~\eqref{EDGC} with a DGC function according to~\eqref{kappadyn} or~\eqref{kappavol}
are a suitable starting point for exploring the effects of DGC.
}} \QEDrem
\end{Remark}

\section{The Mathematical Structure of Dynamical Gravitational Coupling} \label{secmath}
\subsection{Conformal Killing Symmetries and Conservation Laws} \label{seckilling}
In preparation, we recall the notion of a conformal Killing field and explain why this notion is
useful for studying the dynamics of gravitational coupling.
A vector field~$K$ is a {\em{Killing field}} if it satisfies the Killing equation
\beq \label{killing}
\nabla_{(j} K_{k)} = 0 \:,
\eeq
where the brackets denote symmetrization, i.e.\ $\nabla_{(j} K_{k)} \equiv \frac{1}{2} \,\big( \nabla_j K_k + \nabla_k K_j \big)$. A Killing field describes a continuous symmetry of space-time. According to Noether's theorem,
continuous symmetries give rise to conservation laws.
For a geodesic~$\gamma(\tau)$, the corresponding conserved quantity is simply
the inner product~$\la K, \dot{\gamma} \big\ra_{\gamma(\tau)}$, as is verified by the computation
\beq \label{killconserve}
\frac{d}{d\tau} \big\la K, \dot{\gamma}(\tau) \big\ra_{\gamma(\tau)} = \nabla_j K_k \:\dot{\gamma}^j \,\dot{\gamma}^k
= \nabla_{(j} K_{k)} \:\dot{\gamma}^j \,\dot{\gamma}^k = 0\:.
\eeq
If~$\gamma$ is a null geodesic, this conservation law can be
generalized to a so-called {\em{conformal Killing field}}, where~\eqref{killing} is
weakened to the conformal Killing equation
\beq \label{confkill}
\nabla_{(j} K_{k)}(x) = \lambda(x)\, g_{jk}(x) \:.
\eeq
Namely,
\[ \frac{d}{d\tau} \big\la K, \dot{\gamma}(\tau) \big\ra_{\gamma(\tau)}
= \nabla_{(j} K_{k)} \:\dot{\gamma}^j \,\dot{\gamma}^k
= \lambda\, g_{jk} \:\dot{\gamma}^j \,\dot{\gamma}^k = 0 \:. \]
Taking the trace of~\eqref{confkill}, one finds that~$\text{div}(K) = 4 \lambda$, implying that
the conformal Killing equation can be written equivalently as
\[ \nabla_{(j} K_{k)} = \frac{1}{4}\:  g_{jk} \;\text{div}\, K \:. \]

\subsection{Behavior under Conformal Transformations} \label{secconftrans}
The DGC function has a simple behavior under conformal transformations, as we now explain.
We consider a conformal transformation of the metric
\beq \label{gconform}
g \rightarrow \tilde{g} = h\, g
\eeq
with a strictly positive smooth function~$h \in C^\infty_0(\scrM)$.
We assume that the conformal transformation is trivial outside a space-time region~$\Omega \subset \scrM$, i.e.
\[ h|_{\scrM \setminus \Omega} \equiv 1 \:. \]
Null geodesics are invariant under conformal transformations, but the affine parameter changes.
In order to see this in detail, let~$\gamma(\tau)$ be a null geodesic corresponding to the metric~$g$, i.e.
\[ 0 = \nabla_\tau \dot{\gamma}^j = \frac{d}{d\tau} \dot{\gamma}^j + \Gamma^j_{kl} \dot{\gamma}^k \dot{\gamma}^l \:. \]
Under the conformal change, the Christoffel symbols transform as follows,
\begin{align*}
\tilde{\Gamma}^j_{kl} &= \frac{1}{2}\: \tilde{g}^{ja} \,\Big( \partial_k \tilde{g}_{la} + \partial_l \tilde{g}_{ka}
- \partial_a \tilde{g}_{kl} \Big) \\
&= \Gamma^j_{kl} + \frac{1}{2 h}\: g^{ja} \,\Big( (\partial_k h) \:g_{la} + (\partial_l h)\: g_{ka} - (\partial_a h)\: g_{kl}
\Big) \\
&= \Gamma^j_{kl} + \frac{1}{2}\:\Big( (\partial_k \log h) \:\delta^j_l +
(\partial_l \log h)\: \delta^j_k - g^{ja} (\partial_a \log h)\: g_{kl} \Big) \:.
\end{align*}
Thus the equation for a null geodesic~$\tilde{\gamma}(\tau)$ becomes
\begin{align}
0 &= \frac{d}{d\tilde{\tau}} \dot{\tilde{\gamma}}^j + \tilde{\Gamma}^j_{kl} \dot{\tilde{\gamma}}^k \dot{\tilde{\gamma}}^l \notag \\
&= \frac{d}{d\tilde{\tau}} \dot{\tilde{\gamma}}^j + \Gamma^j_{kl} \dot{\tilde{\gamma}}^k \dot{\tilde{\gamma}}^l
+ (\partial_k \log h) \:\dot{\tilde{\gamma}}^k \dot{\tilde{\gamma}}^j \label{geodesic}
\end{align}
(where in the last step we used that~$g_{kl} \dot{\tilde{\gamma}}^k \dot{\tilde{\gamma}}^l=0$).
Reparametrizing the original null geodesic by setting
\beq \label{geodef}
\tilde{\gamma}(\tilde{\tau}) = \gamma(\tau) \qquad \text{with} \qquad
\frac{d\tilde{\tau}}{d\tau} = h\big(\gamma(\tau) \big) \:,
\eeq
we find that
\begin{align*}
\dot{\tilde{\gamma}}(\tilde{\tau}) &= \frac{d}{d\tilde{\tau}} \tilde{\gamma}(\tilde{\tau})
=  \frac{d\tau}{d\tilde{\tau}}\: \frac{d}{d\tau} \gamma(\tau) = \frac{1}{h(\gamma(\tau))} \: \dot{\gamma}(\tau) \\
\frac{d}{d\tilde{\tau}} \dot{\tilde{\gamma}}^j &= \frac{1}{h(\gamma(\tau))}
\frac{d}{d\tau} \bigg( \frac{1}{h(\gamma(\tau))} \: \dot{\gamma}^j(\tau) \bigg) \\
&= \frac{1}{h^2}\: \frac{d}{d\tau} \dot{\gamma}^j(\tau) -
\frac{1}{h^3}\: \Big(\partial_k h\big(\gamma(\tau)\big) \Big) \:\dot{\gamma}^k(\tau) \:\dot{\gamma}^j(\tau) \\
&= \frac{1}{h^2}\: \frac{d}{d\tau} \dot{\gamma}^j -
\frac{1}{h^2}\: (\partial_k \log h ) \:\dot{\gamma}^k \dot{\gamma}^j \\
\Gamma^j_{kl} \dot{\tilde{\gamma}}^k \dot{\tilde{\gamma}}^l &= \frac{1}{h^2}\:
\Gamma^j_{kl} \dot{\gamma}^k \dot{\gamma}^l \\
(\partial_k \log h) \:\dot{\tilde{\gamma}}^k \dot{\tilde{\gamma}}^j &=
\frac{1}{h^2}\: (\partial_k \log h) \:\dot{\gamma}^k \dot{\gamma}^j \:.
\end{align*}
Using these formulas, one sees that~$\tilde{\gamma}$ defined by~\eqref{geodef}
indeed satisfies~\eqref{geodesic} and is thus a null geodesic.

Using this transformation law for null geodesics under conformal transformations, one
finds that
\begin{align}
\dot{\tilde{\gamma}} &= \frac{1}{h}\: \dot{\gamma} \label{conf1} \\
D_p\tilde{\scrL} &= \Big\{ \frac{u}{h(p)} \:\Big|\: u \in D_p\scrL \Big\} \\
\tilde{\mu}_p \big( D_p \tilde{\scrL} \big) &= h(p) \, \mu_p \big( D_p \tilde{\scrL} \big)
= \frac{1}{h(p)} \:\mu_p \big( D_p\scrL \big) \\
\tilde{\kappa}_\vol(p) &= h(p)\: \kappa_\vol(p) \qquad \text{and} \qquad
\tilde{\kappa}(p) = h(p)\: \kappa(p) \:. \label{conf4}
\end{align}
The last formula is what one would have expected from a naive scaling argument,
keeping in mind that~$\kappa$ has the dimension of length squared.

\subsection{Local Existence and Uniqueness of the Time Evolution} \label{seccauchy}
In this section, we prove local existence and uniqueness of the time evolution
for the Einstein equations with DGC.
Before beginning, we recall that without matter, the Einstein equations with
DGC coincide with the usual vacuum Einstein equations~$R_{jk}=0$ 
(see the paragraph after~\eqref{Etrace}).
This shows in particular that, starting with smooth initial data, singularities may form in finite
time (see~\cite{christoBH, li-yu}). This is why we restrict attention to {\em{local}} solutions
to the Cauchy problem. In preparation for the Cauchy problem for the whole system, let us consider the
Cauchy problem for our DGC models in a given space-time background.
In order to pose the Cauchy problem, one must assume that
the space-time is {\em{globally hyperbolic}}
(for basic definitions see for example~\cite{baer+ginoux, choquet, ringstroem}). Let~$\scrN$ be a Cauchy surface.
Then, by definition of global hyperbolicity, every maximal null geodesic intersects~$\scrN$ exactly once.
Therefore, the set~$\scrL$, \eqref{Gammadef}, is characterized uniquely by looking at the
vectors~$\dot{\gamma}$ on~$\scrN$,
\[ D_\scrN\scrL := \bigcup_{p \in \scrN} D_p\scrL
= \Big\{ \dot{\gamma}(\tau) \:\Big|\: \text{$(\gamma, I) \in \scrL$, $\tau \in I$
and $\gamma(\tau) \in \scrN$} \Big\} \:. \]
Conversely, initial-data of this form gives rise to a unique family of null geodesics~$\scrL$,
as is made precise in the following proposition.

\begin{Prp} \label{prpG}
Let~$\mathscr{G}_\scrN$ be a set of null vectors
\[ \mathscr{G}_\scrN = \bigcup_{p \in \scrN} \mathscr{G}_p \qquad \text{with} \qquad
\mathscr{G}_p = \Big\{ v_p \in T_pM \:\Big|\: v_p \neq 0 \text{ and } \la v_p, v_p \ra_p = 0 \Big\} \]
with the following properties:
\begin{itemize}[leftmargin=2em]
\item[(a)] If~$v_p$ is in~${\mathscr{G}}_p$, so is~$-v_p$.
\item[(b)] For every~$p \in \scrN$ and for every non-zero null vectors~$u \in T_pM$,
there is a unique scalar~$\lambda >0$ such that~$\lambda u \in \mathscr{G}_\scrN$.
\end{itemize}
Then there is a unique choice of parametrized null geodesics~$\scrL$
(compatible with Definition~\ref{defL}) such that~$D_\scrN\scrL = \mathscr{G}$.

If the sets~$\mathscr{G}_p$ are smooth surfaces for each~$p \in \scrN$ and depend smoothly
on the base point~$p \in \scrN$, then the sets~$D_p\scrL$ are smooth surfaces for all~$p \in \scrM$.
\end{Prp}
\Proof In order to construct~$\scrL$, one solves the geodesic equation with initial values on~$\scrN$,
\[ \nabla_\tau \gamma(\tau) = 0 \:,\qquad \gamma(\tau) = p \:,\quad \dot{\gamma}(\tau_0) = \pm v_p\:. \]
Taking all the resulting maximal geodesics for all choices of~$\tau_0 \in \R$ and
all~$v_p \in \scrL$ gives a family~$\scrL$.
The uniqueness of~$\scrL$ follows immediately from the fact that every maximal null geodesic
intersects~$\scrN$, where the parametrization is determined by~$\mathscr{G}_\scrN$.

The smoothness is an immediate consequence of the fact that the solutions of ordinary
differential equations depend smoothly on parameters and on the initial data.
\QED

We next consider the time evolution of the DGC tensor, in a given globally hyperbolic space-time
with given~$\scrL$. According to~\eqref{Edef}, it can be computed at a space-time point~$p$
if for every future-directed~$x \in D_p\scrL$ we know the tangent vector
\beq \label{Jint}
J^l(p,x) :=
\int_{\tau_{\min}}^0 d\tau \:\big(\Pi_{p, \exp_p(\tau x)} \big)^{la}
\: \Big( \ell^{-1}_{ab} \:\big(\partial_c \kappa \big) \: T^{bc} \Big) \Big|_{\exp_p(\tau x)}\:.
\eeq
In words, this tangent vector is obtained by integrating along a null geodesic to the past.
Since every null geodesic intersects the Cauchy surface~$\scrN$, one can describe~$J$
uniquely by prescribing it on on the Cauchy surface as a mapping
\beq \label{Jinit}
J|_\scrN \::\: D_\scrN\scrL^\vee \rightarrow T\scrM \qquad \text{with} \qquad
J|_\scrN(p,x) \in T_p\scrM \quad \text{for all~$(p, x) \in D_\scrN\scrL^\vee_\scrN$}\:,
\eeq
where the~$\scrL^\vee$ denotes the set of all future-directed null geodesics in~$\scrL$.
Then the time evolution of~$J$ is obtained by solving the transport equation
\beq \label{Jdyn}
\nabla_x J(p,x) = \ell^{-1}_{ab}(p) \:\big(\partial_c \kappa(p) \big) \: T^{bc}(p) \:,
\eeq
giving a mapping~$J : D\scrL^\vee \rightarrow T\scrM$
(where~$D\scrL^\vee = \cup_{p \in \scrM} D_p \scrL^\vee$ with the obvious fiber bundle structure).
In principle, the mapping~$J|_\scrN$ should be determined according to~\eqref{Jint} and~\eqref{Edef}
by integrating along and over all null geodesics to the past.
In situations in which this procedure is not feasible (for example if the past development of the Cauchy
surface is not known), one should regard~$J|_\scrN$ as part of the initial data,
which should be chosen depending on the physical situation in mind.
Whenever the effect of the DGC tensor is small, one may simply choose~$J|_\scrN \equiv 0$.

We now turn our attention to the Cauchy problem for our DGC models.
Assume that we are given a Cauchy surface~$\scrN$.
On~$\scrN$ we are given a smooth Riemannian metric~$\overline{g}$
and a smooth second fundamental form~$h$. Due to finite propagation speed, it suffices to solve the Cauchy
problem in a small open subset of~$\scrN$ (then the solution on~$\scrN$ is obtained by ``glueing together'' the local solutions). With this in mind, we may work in a local chart and identify~$\scrN$ with a subset of~$\R^3$.
Working in the wave gauge, the Ricci tensor is obtained from the metric by applying
a quasilinear hyperbolic operator (for details see~\cite{foures-bruhat, choquet}),
\[ R_{jk} = -\frac{1}{2}\: g^{ab} \partial_{ab} g_{jk} + \text{(l.o.t.)}\:. \]
Moreover, the Riemannian metric and second fundamental form on~$\scrN$ give rise to
the Lorentzian metric in the wave gauge and its first time derivative (for details see~\cite{choquet}).
Therefore, we may describe the initial data of the gravitational field by
\[ g_{jk}|_{t=0} \qquad \text{and} \qquad \partial_t g_{jk}|_{t=0} \:. \]
The matter is described by additional fields on~$\scrN$.
We assume that the equations of matter are such that coupling the matter fields to the
usual Einstein equations, one obtains a system of partial differential equations for which the
Cauchy problem is well-posed (like for example a perfect fluid, Vlasov matter, the Maxwell field,
the Yang-Mills field, a scalar field, the Dirac field or the Klein-Gordon field).
We also assume that our initial data satisfies the Einstein {\em{constraint equations}}
(see for example~\cite[Section~III.13]{ringstroem} or~\cite[Section~VII]{choquet}).
This problem can be treated exactly as for the Einstein's gravity.
Particular solutions of the constraint equations are obtained by taking
initial data of the usual Einstein equations and choosing DGC initial data with~$\kappa|_{\scrN} \equiv 8 \pi G$
and~$J|_{\scrN} \equiv 0$. But clearly, choices with non-trivial DGC on~$\scrN$
are also possible.

Under these assumptions we prove the following result:
\begin{Thm} \label{thmcauchy}
The Cauchy problem for the Einstein-matter equations with DGC
is well-posed.
\end{Thm}
The remainder of this section is devoted to the proof of this theorem.
In more technical terms, our above assumptions on the matter fields can be
restated by saying that the coupled Einstein-matter system can be rewritten
as a quasilinear symmetric hyperbolic system (see~\cite{ringstroem}). 
In order to show that the Cauchy problem remains well-posed if DGC
is taken into account, we need to settle the following issues:
\begin{itemize}[leftmargin=2.5em]
\item[(a)] Given the gravitational coupling function~$\kappa(x)$, the 
modified Einstein-matter equations can again be written as a quasilinear symmetric hyperbolic system.
\item[(b)] The dynamics of the DGC function~$\kappa(x)$
(as defined by~\eqref{kappadyn} or~\eqref{kappavol}) 
and of the DGC tensor~$E_{jk}$ (as defined by~\eqref{Edef} in Appendix~\ref{appE})
can be described by a quasilinear symmetric hyperbolic system.

\end{itemize}
Once these points are proven, by combining the symmetric hyperbolic systems
of the Einstein-matter system and of the dynamical equations for~$\kappa(t,x)$
and~$E_{jk}(t,x)$,
one obtains a symmetric hyperbolic system for the Einstein-matter equations
with DGC. Then Theorem~\ref{thmcauchy} follows
from the general local existence result for solutions of the initial value
problem for quasilinear symmetric hyperbolic systems (see~\cite[Section~9]{ringstroem}
or~\cite[Section~16]{taylor3}).

The next consideration settles the above problem~(a):
The highest orders of the derivatives of the metric do not involve derivatives of~$\kappa(x)$. Thus the resulting equations
are of the form
\[ R_{jk} - \frac{1}{2}\: R\, g_{jk} = \kappa(x) \,T_{jk} + \text{(l.o.t)} \:. \]
Since the lower order terms only affect the zero order terms in the corresponding first order
system, the system is again symmetric hyperbolic.

We next explain how to settle the above problem~(b). Our task is to
write the dynamics of~$\kappa(x)$ and~$E_{jk}(x)$ in terms of a quasilinear symmetric hyperbolic system.
To this end, it is convenient to describe the initial data~$\mathscr{G}_\scrN$
in Proposition~\ref{prpG} by a smooth mapping
\beq \label{sigmainit}
\sigma \::\: \scrN \times S^2 \rightarrow \R^+ \:.
\eeq
Given~$\sigma$, we construct the set~$D_p \scrL$ for any~$p \in \scrN$ as follows.
To any~$u \in S^2 \subset \R^3$ we associate a null vector~$n_p \in T_p\scrN$ by
\[ n_p = n^0_p \:\frac{\partial}{\partial t} \Big|_p
+ \sum_{\alpha=1}^3 u^\alpha \: \frac{\partial}{\partial x^\alpha} \Big|_p \:, \]
where the time component~$n^0>0$ is determined uniquely by imposing that~$\la n_p, n_p\ra_p =0$.
Now we set
\[ D_p \scrL = \big\{ \sigma \big(p,u \big)\, n_p \:\big|\: n \in S^2 \big\} . \]

In order to describe the dynamics of the function~$\sigma$, we follow the flow of
the null geodesics. Thus given~$(p, u) \in \scrN \times S^2$, we let~$\gamma(\tau)$ be the
null geodesic with initial conditions
\[ \gamma(0)=p \qquad \text{and} \qquad \dot{\gamma}(0) = n_p \]
(and the dot again denotes the derivative with respect to~$\tau$). Thus in components,
\beq \label{dotg}
\dot{\gamma}(0) = \sigma\big(p,u\big) \begin{pmatrix} n^0 \\ u \end{pmatrix} \:.
\eeq
Using the geodesic equation
\[ \frac{d}{d\tau} \dot{\gamma}^j = -\Gamma^j_{kl} \dot{\gamma}^k \dot{\gamma}^l \:, \]
the position and the tangent vector of the geodesic have evolved to
\begin{align*}
p^\alpha(t) &= p^\alpha + t\, \frac{\dot{\gamma}^\alpha(0)}{\dot{\gamma}^0(0)} + \O\big(t^2)= 
p^\alpha + \frac{t}{n^0}\:  u^\alpha + \O\big(t^2) \\
\dot{\gamma}^\alpha(t) &= \dot{\gamma}^\alpha(0)
- t \:\Gamma^\alpha_{kl} \big|_p \,\dot{\gamma}^k(0)\, \dot{\gamma}^l(0) + \O\big(t^2) \\
&= \sigma\big(p,u\big) \, u^\alpha - t \:\Gamma^\alpha_{kl} \big|_p \,\dot{\gamma}^k(0)\, \dot{\gamma}^l(0)
+ \O\big(t^2) \:.
\end{align*}
(here~$\alpha=1,2,3$ denotes a spatial tensor index). Decomposing the
vector~$\Gamma^\alpha_{kl} \big|_p \,\dot{\gamma}^k\, \dot{\gamma}^l$ into
the components which are parallel and orthogonal to~$u$, we write the last equation as
\[ \dot{\gamma}^\alpha(t) = \sigma\big(p,u\big) \, u^\alpha + t \, \big( v_\perp^\alpha
+ v^\alpha_\parallel \big) + \O\big(t^2) \:, \]
where
\beq \label{vppdef}
v_\parallel := -u \, \sum_{\alpha=1}^3 u^\alpha \Gamma^\alpha_{kl} \big|_p \,\dot{\gamma}^k\, \dot{\gamma}^l
\qquad \text{and} \qquad 
v_\perp := -\Gamma^\alpha_{kl} \big|_p \,\dot{\gamma}^k\, \dot{\gamma}^l - v_\parallel\:.
\eeq
We thus obtain
\begin{align*}
\sigma\bigg(t, p + \frac{t}{n^0}\:  u,  u + t \,\frac{v_\perp}{\sigma(p,u)} \bigg)
&= \sigma(p,u) + t \,\|v_\parallel\|_{\R^3} + \O\big(t^2)
\end{align*}
(note that the vector~$u + t \,v_\perp$ is again a unit vector up to errors of the order~$\O\big(t^2)$).
Dividing by~$t$ and taking the limit~$t \rightarrow 0$, we obtain the differential equation
\beq \label{deq}
\partial_t \sigma|_{(t, p,u)} + \frac{1}{n^0}\, D_2 \sigma|_{(t, p,u)} \,u
+ \frac{1}{\sigma(t,p,u)}\: D_3 \sigma|_{(t, p,u)} \,v_\perp = \|v_\parallel\|_{\R^3}
\eeq
(here~$D_2$ and~$D_3$ denote the partial derivatives with respect to the second and
and third arguments, and applying the vector gives the corresponding directional derivative).
Since all the coefficients are real-valued and the coefficient of~$\partial_t \sigma$ is positive,
this is obviously a symmetric hyperbolic system. In order to determine the coefficients of
this system, one first defines~$\dot{\gamma}(t,p,u)$ similar to~\eqref{dotg} by
\[ \dot{\gamma}(t,p,u) = \sigma\big(t, p,u\big) \begin{pmatrix} n^0(t,p,u) \\ u \end{pmatrix} \]
and computes~$n^0(t,p,u)$ as the unique solution of the algebraic equation
\[ g(t,p)_{jk} \,\dot{\gamma}^j(t,p,u)\, \dot{\gamma}^j(t,p,u) = 0 \:. \]
The vectors~$v_\parallel(t,p,u)$ and~$v_\perp(t,p,u)$ are then defined by~\eqref{vppdef}.
In this way, the coefficients are obtained by solving algebraic equations.
This shows that the system is indeed quasilinear and symmetric hyperbolic.

Since the DGC tensor can be obtained from the vector field~$J$ in~\eqref{Jint} by
integrating over~$D_p\scrL$ (see~\eqref{Edef}), its dynamics is described completely
by that of~$J$. The initial data~\eqref{Jinit} can be written in local coordinates
similar to~\eqref{sigmainit} by a mapping~$J|_{\scrN} : \scrN \times S^2 \rightarrow \R^4$,
where the vector in~$R^4$ describes the components of the tangent space in the considered chart.
Writing the transport equations~\eqref{Jdyn} in our chart gives a linear hyperbolic system.

This concludes the proof for the simplified model~\eqref{kappavol}.
For the DGC function~\eqref{kappadyn}, there is the complication that
computing the Gaussian curvature~$K_p$ makes it necessary to
compute the second derivatives of~$\sigma$ with respect to the variable~$u$.
In order to treat these derivatives, one differentiates~\eqref{deq} to obtain additional
differential equations. For example
\[ \partial_t D_3 \sigma|_{(t, p,u)} + \frac{1}{n^0}\, D_2 D_3 \sigma|_{(t, p,u)} \,u
+ \frac{1}{\sigma(t,p,u)}\: D_3 D_3 \sigma|_{(t, p,u)} \,v_\perp = D_3 \|v_\parallel\|_{\R^3} 
+ \text{(l.o.t)}\:, \]
and similar for the second derivatives.
Including these additional differential equations to the symmetric hyperbolic system,
we again obtain a quasilinear symmetric hyperbolic system.
Note that, differentiating the initial data~$\sigma(p,u)$ with respect to~$u$, one also gets corresponding
initial data for~$D_3 \sigma|_{(0, p,u)}$ and~$D_3^2 \sigma|_{(0, p,u)}$.
This concludes the proof of Theorem~\ref{thmcauchy}.

\section{Example: Modifications to the Friedmann-Robertson-Walker Model} \label{secfriedmann}
As a concrete example, we now consider the metric of the Friedmann-Robert\-son-Walker (FRW) geometry
\beq \label{frw}
ds^2 = dt^2 - R^2(t) \,d\sigma^2 \:,
\eeq
where~$t$ is the time for a co-moving observer, $R$ is the scale function, and~$d\sigma$ is the
line element on the unit three-sphere, the flat~$\R^3$ or a three-dimensional hyperboloid.
The line element~\eqref{frw} is spatially homogeneous and isotropic.
Therefore, it is natural to assume that the set of parametrized null geodesics~$\scrL$
is also homogeneous and isotropic. This implies that the sets~$D_p\scrL$ are of the form
\beq \label{LFRW}
D_{(t,x)}\scrL = \left\{ \pm \frac{1}{\varepsilon(t)}\: \big( 1, \vec{n} \big) \quad \text{with} \quad
\sigma(\vec{n}, \vec{n}) = 1  \right\} .
\eeq
In this way, the DGC function is described by a single function~$\varepsilon(t)$.
In order to compute this function, we introduce the new time function~$\tau$ by
\[ \frac{d\tau}{dt} = R \:. \]
Then the line element takes the form
\[ ds^2 = R^2 \,\big( d\tau^2 - d\sigma^2 \big) \:. \]
This metric is conformal to the static metric~$d\tau^2 - d\sigma^2$.
For this static metric, the vector field~$K:=\partial_\tau$ is a Killing field,
implying that the inner product~\eqref{killconserve} is conserved.
As a consequence, the function~$\varepsilon(t) =\varepsilon$ is constant for the static metric.
Comparing with the behavior under conformal transformations~\eqref{gconform} and~\eqref{conf1},
we conclude that~$h(\tau) = R^2(\tau)$ and~$\varepsilon(\tau) \sim R(\tau)$.
Using~\eqref{conf4}, we conclude that
\[ \kappa(t) = R^2(t)\, \kappa \]
with a constant~$\kappa$.
Thus the effective gravitational constant grows quadratically in the scale function.
We remark that this is quite similar to the behavior in Brans-Dicke theory as found in~\cite{dehnen}.
However, our result is different from Dirac's proposal in~\cite[Section~5]{dirac4},
who conjectured that the gravitational constant should decrease in time.

Taking the FRW metric as a model for our universe, the above findings
imply that in the early universe, the effective gravitational constant should have been smaller.
The resulting effect seems similar to the expansion of the early universe as proposed by {\em{inflationary
cosmological models}}. At present, it is unknown if and to what extend DGC can reproduce the
results of inflation.

The FRW and inflationary models also lead to the following
general {\em{method for prescribing~$\scrL$}}:
Since the early universe seems to have been in a highly homogeneous
and isotropic configuration (corresponding to the symmetries of the FRW model),
and keeping in mind that in the FRW model there is a canonical choice of~$\scrL$, it seems natural to
make the physical assumption that~$\scrL$ should be chosen such that in the early universe it
agrees with~\eqref{LFRW}. This prescription involves only one free parameter~$\varepsilon$
(which determines the gravitational constant) and thus makes DGC to a model of our universe
with predictive power. But clearly, this procedure involves the difficulty
that in order to make predictions at present time, one must know the time evolution of~$\scrL$
starting from the early universe.
From the theoretical point of view, it seems interesting that this procedure
{\em{distinguishes the future from the past}}, because~$\scrL$ is determined by initial values
right after the big bang. This point of view might even make it possible to define the arrow of time
and notions like {\em{entropy}} via the ``increasing disorder'' of the null geodesics in~$\scrL$ as time evolves.

\section{Example: The Spherically Symmetric Collapse} \label{seccollapse}
In this section, we explore DGC in the gravitational collapse of a
spherically symmetric star.
In order to work in a clean and explicit example, we consider a spherically symmetric
shell of matter, i.e.\ the situation that all the matter of the star is concentrated on its surface.
Despite its simplicity, this model of a {\em{collapsing shell of matter}}
captures all relevant effects of dynamical gravitational coupling.
This is because our analysis as well as our results generalize in a straightforward way to models with
several shells. By taking the number of shells sufficiently large, one can approximate a
realistic spherically symmetric star with an arbitrary mass density.

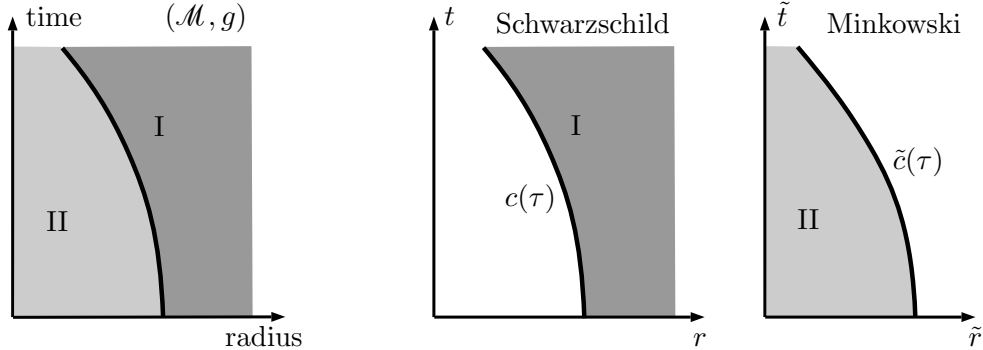
\begin{figure}
\psscalebox{1.0 1.0} 
{
\begin{pspicture}(0,-3.4190235)(16.82,3.4190235)
\definecolor{colour0}{rgb}{0.8,0.8,0.8}
\definecolor{colour1}{rgb}{0.6,0.6,0.6}
\pspolygon[linecolor=white, linewidth=0.02, fillstyle=solid,fillcolor=colour0](10.03,-0.72902346)(12.015,-0.72402346)(11.995,-0.36902344)(11.95,0.115976565)(11.86,0.57097656)(11.665,1.0909766)(11.37,1.6409765)(11.13,2.0409765)(10.815,2.4759765)(10.475,2.8809767)(10.01,2.8859766)
\pspolygon[linecolor=white, linewidth=0.02, fillstyle=solid,fillcolor=colour1](7.6282907,-0.7290027)(8.84,-0.72902346)(8.825,2.885935)(6.275,2.8759766)(6.5477176,2.5043583)(6.7749405,2.1679683)(6.9668183,1.8416197)(7.1334486,1.4901674)(7.234437,1.2441509)(7.335425,0.983072)(7.4465117,0.62659895)(7.522253,0.24000145)(7.5676975,-0.056222606)(7.6030436,-0.50808984)
\pspolygon[linecolor=white, linewidth=0.02, fillstyle=solid,fillcolor=colour1](2.0082905,-0.7290027)(3.22,-0.72902346)(3.205,2.885935)(0.655,2.8759766)(0.9277174,2.5043583)(1.1549407,2.1679683)(1.3468182,1.8416197)(1.5134486,1.4901674)(1.6144367,1.2441509)(1.7154249,0.983072)(1.8265119,0.62659895)(1.9022529,0.24000145)(1.9476976,-0.056222606)(1.9830434,-0.50808984)
\pspolygon[linecolor=white, linewidth=0.02, fillstyle=solid,fillcolor=colour0](0.035,-0.73402345)(2.02,-0.72902346)(2.0,-0.37402344)(1.955,0.11097656)(1.865,0.56597656)(1.705,1.0959766)(1.475,1.6359766)(1.27,2.0409765)(0.97,2.4959764)(0.665,2.8759766)(0.015,2.8809767)
\rput[bl](0.195,3.1459765){\normalsize{time}}
\psframe[linecolor=white, linewidth=0.02, dimen=outer](1.01,-3.3790236)(0.97,-3.4190235)
\psline[linecolor=black, linewidth=0.04, arrowsize=0.05291666666666668cm 2.0,arrowlength=1.4,arrowinset=0.0]{->}(0.02,-0.72902346)(0.02,3.2709765)
\rput[bl](2.845,-1.1140234){\normalsize{radius}}
\psline[linecolor=black, linewidth=0.04, arrowsize=0.05291666666666668cm 2.0,arrowlength=1.4,arrowinset=0.0]{->}(0.025002837,-0.73204625)(3.6449971,-0.7260006)
\psline[linecolor=black, linewidth=0.04, arrowsize=0.05291666666666668cm 2.0,arrowlength=1.4,arrowinset=0.0]{->}(5.62,-0.72902346)(5.62,3.2709765)
\psline[linecolor=black, linewidth=0.04, arrowsize=0.05291666666666668cm 2.0,arrowlength=1.4,arrowinset=0.0]{->}(5.625,-0.72902346)(9.22,-0.72902346)
\psline[linecolor=black, linewidth=0.04, arrowsize=0.05291666666666668cm 2.0,arrowlength=1.4,arrowinset=0.0]{->}(10.025,-0.72902346)(12.82,-0.72902346)
\psline[linecolor=black, linewidth=0.04, arrowsize=0.05291666666666668cm 2.0,arrowlength=1.4,arrowinset=0.0]{->}(10.02,-0.72902346)(10.02,3.2709765)
\psbezier[linecolor=black, linewidth=0.06](2.02,-0.73902345)(1.9821527,0.35225093)(1.8037274,0.89545465)(1.5586885,1.4539616)(1.3136497,2.0124686)(1.0537225,2.4069571)(0.68,2.8509765)
\psbezier[linecolor=black, linewidth=0.06](12.02,-0.72402346)(11.982153,0.36725092)(11.753727,0.96545464)(11.433688,1.5139617)(11.113649,2.0624685)(10.828722,2.4219573)(10.455,2.8659766)
\psbezier[linecolor=black, linewidth=0.06](7.62,-0.74402344)(7.582153,0.34725094)(7.4037275,0.90045464)(7.1586885,1.4589616)(6.9136496,2.0174687)(6.6537223,2.4119573)(6.28,2.8559766)
\rput[bl](1.9,1.6859765){\normalsize{I}}
\rput[bl](0.47,0.41597655){\normalsize{II}}
\rput[bl](7.44,1.6909766){\normalsize{I}}
\rput[bl](5.76,3.1609766){\normalsize{$t$}}
\rput[bl](9.065,-1.1090235){\normalsize{$r$}}
\rput[bl](6.565,0.65597653){\normalsize{$c(\tau)$}}
\rput[bl](10.455,0.44597656){\normalsize{II}}
\rput[bl](10.185,3.1409767){\normalsize{$\tilde{t}$}}
\rput[bl](12.725,-1.1190234){\normalsize{$\tilde{r}$}}
\rput[bl](11.72,1.1409765){\normalsize{$\tilde{c}(\tau)$}}
\rput[bl](2.045,3.0559766){\normalsize{$(\scrM, g)$}}
\rput[bl](6.445,3.0559766){\normalsize{Schwarzschild}}
\rput[bl](10.845,3.0559766){\normalsize{Minkowski}}
\end{pspicture}
}
\caption{A collapsing shell of matter.}
\label{figcollapse}
\end{figure}

The space-time~$(\scrM, g)$ we have in mind is shown on the left of Figure~\ref{figcollapse},
where the shell of matter is plotted as a function of a time and a radial variable
(as usual, the angular variables~$\vartheta$ and~$\varphi$
are not shown, so that every point in the figure corresponds to
a two-dimensional sphere).
Since all the matter is concentrated on the shell, both in the exterior and interior
space-time regions (denoted in the figure by~I respectively~II) the vacuum Einstein equations hold.
Due to Birkhoff's theorem, these space-time regions are isometric to subsets of Schwarzschild space-time.
Thus region~I is isometric to a subset of Schwarzschild space-time of mass~$M$, whereas
region~II is isometric to a subset of Minkowski space.
For convenience, in region~I we choose Schwarzschild coordinates~$(t,r)$, i.e.
\[ ds^2 = \left( 1 - \frac{2M}{r} \right) dt^2 - \left( 1 - \frac{2M}{r} \right)^{-1} dr^2 
- r^2 \big( d\vartheta^2 + \cos^2 \vartheta\: d\varphi^2 \big) \:. \]
Likewise, in region~II we choose polar coordinates~$(\tilde{t}, \tilde{r}, \vartheta, \varphi)$, i.e.
\[ ds^2 = d{\tilde{t}}^{\,2} - d\tilde{r}^2 - \tilde{r}^2 \big( d\vartheta^2 + \cos^2 \vartheta\: d\varphi^2 \big) \]
(see the middle and right plots in Figure~\ref{figcollapse}).
The space-time~$(\scrM, g)$ is obtained by ``glueing together'' these subsets of Schwarzschild and
Minkowski space along their boundaries. Due to spherical symmetry, we may disregard
the angular variables. Then the boundaries of these space-time regions can be
described by timelike curves~$c(\tau)$ and~$\tilde{c}(\tau)$,
which both have no angular components, i.e.
\beq \label{ctau}
c(\tau) = \big( c^0(\tau), c^1(\tau), 0, 0 \big) \qquad \text{and} \qquad
\tilde{c}(\tau) = \big( \tilde{c}^{\,0}(\tau), \tilde{c}^1(\tau), 0, 0 \big) \:.
\eeq

We next explain the glueing and specify the choice of the curves~$c$ and~$\tilde{c}$.
For simplicity, we begin with the situation that the outer surface of shell is freely falling.
Therefore, we choose~$c(\tau)$
as a timelike geodesic, which we parametrize by arc length, i.e.
\[ \nabla_\tau \dot{c}(\tau)= 0 \quad \text{and} \quad
\big\la \dot{c}(\tau), \dot{c}(\tau) \big\ra = 1 \qquad \text{for all~$\tau$} \]
(where~$\la .,.\ra$ is the Schwarzschild metric of mass~$M$).
We want to choose the curve~$\tilde{c}(\tau)$ in such a way that the ``glueing'' can be accomplished simply by
\[ \text{identifying~$c(\tau)$ with~$\tilde{c}(\tau)$ for all~$\tau$.} \]
We must make sure that the Lorentzian metric is continuous across the glueing surface.
It suffices to consider the tangential component of the metric, because the transversal
components can be arranged to be continuous by choosing Fermi coordinates in a tubular neighborhood
of the glueing surface.
Since the angular components of the metric involve factors~$r^2$ and~$\tilde{r}^2$, respectively,
one condition is that the glueing must be performed for the same radii, i.e.
\beq \label{glue1}
c^1(\tau) = \tilde{c}^1(\tau) \qquad \text{for all~$\tau$} \:.
\eeq
Moreover, the continuity of the metric along the curves~$c$ and~$\tilde{c}$ implies that
also~$\tilde{c}$ must be parametrized by arc length, i.e.
\[ \big\la \dot{\tilde{c}}(\tau), \dot{\tilde{c}}(\tau) \big\ra = 1 \qquad \text{for all~$\tau$} \]
(where~$\la .,. \ra$ is the Minkowski metric).
Solving this equation for the time component and using~\eqref{glue1} gives
\[ \dot{\tilde{c}}^{\,0}(\tau) =  \sqrt{1 + \big(\dot{\tilde{c}}^1(\tau) \big)^2} =
\sqrt{1 + \big(\dot{c}^1(\tau) \big)^2} \:, \]
where in the last step we substituted the $\tau$-derivative of~\eqref{glue1}.
Integration gives
\beq \label{glue2}
\tilde{c}^{\,0}(\tau) = \int^\tau \sqrt{1 + \big(\dot{c}^1(\tau') \big)^2}\: d\tau' + \text{const} \:.
\eeq
With~\eqref{glue1} and~\eqref{glue2} we have computed the curve~$\tilde{c}(\tau)$,
which is unique up to irrelevant time translations.
We thus obtain the desired space-time~$(\scrM, g)$.

Our goal is to determine the strength of the gravitational coupling at a space-time point~$p$.
In order to keep the computations as simple as possible, we restrict attention to the
simplified model with the DGC function~$\kappa_\vol$ introduced in~\eqref{kappavol}
(in the last paragraph of this section, we remark how our results carry over to
the DGC function~\eqref{kappadyn}).
In order to determine the affine parametrizations of all null geodesics through~$p$,
we follow the geodesics to the past until they reach the asymptotic end
(the reason why go to the past and not to the future
corresponds to the general prescription for determining~$\scrL$
explained in the last paragraph of Section~\ref{secfriedmann}).
In the asymptotic end, we then choose the parametrization
as in the example of Minkowski space in Section~\ref{exmink}.
In order to work out the dependence on the velocity vector~$v$ in~\eqref{vdef} and~\eqref{LMinkv},
we first consider the case~$v=(1,\vec{0})$ where the star is at rest in the reference frame
distinguished by the regularization. In this case, we can use~\eqref{LMink} to determine the
parametrization of a future-directed null geodesic~$\gamma(\tau)$ by the condition
\beq \label{viszero}
\lim_{\tau \rightarrow -\infty} \dot{\gamma}^0(\tau) = \frac{1}{\varepsilon} \:.
\eeq
At the end of this section, we shall consider the case of general~$v$ and discuss
how the picture changes if the star moves relative to the reference frame
distinguished by the regularization.

In preparation, we analyze the angular dependence of a null geodesic~$\gamma(\tau)$.
Using the spherical symmetry of the metric, we can assume without loss of generality that
the point~$p$ has angular coordinates~$\vartheta=\frac{\pi}{2}$ and~$\varphi=0$.
Moreover, by a rotation about the axis going through the center of the star and the point~$p$,
we can arrange that the geodesic~$\gamma$ lies in the equatorial plane~$\{ \vartheta = \frac{\pi}{2}\}$.
Next, we can make use of the fact that the vector field~$L := \partial_\varphi$ is a Killing field. According
to~\eqref{killconserve}, it gives rise to the conservation law~$\la L, \dot{\gamma} \ra=-\ell$, where~$\ell$
denotes the angular momentum of the geodesic. Thus the angular dependence is described by the equations
\[ \dot{\gamma}^2(\tau) = 0 \qquad \text{and} \qquad
\dot{\gamma}^3(\tau) = \left\{ \begin{array}{cl} \displaystyle -\frac{\ell}{r(\tau)^2} & \text{in region~I} \\[.8em]
\displaystyle -\frac{\ell}{\tilde{r}(\tau)^2} & \text{in region~II}\:.
\end{array} \right. \]

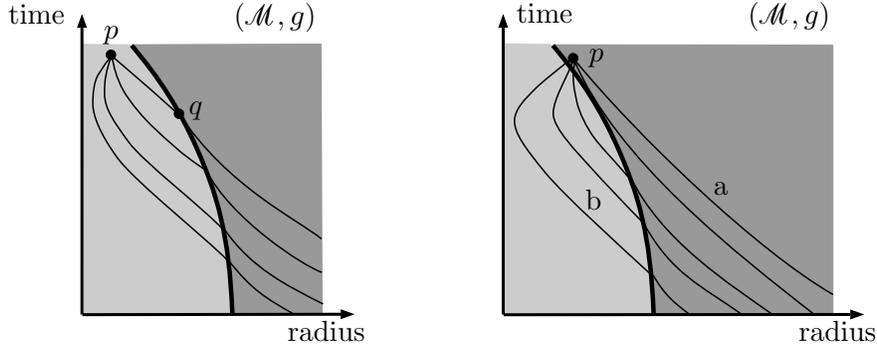
\begin{figure}
\psscalebox{1.0 1.0} 
{
\begin{pspicture}(0,-3.4118261)(14.295,3.4118261)
\definecolor{colour0}{rgb}{0.6,0.6,0.6}
\definecolor{colour1}{rgb}{0.8,0.8,0.8}
\pspolygon[linecolor=white, linewidth=0.02, fillstyle=solid,fillcolor=colour0](2.9782906,-0.72180545)(4.19,-0.7218262)(4.175,2.8931324)(1.625,2.883174)(1.8977174,2.5115557)(2.1249406,2.1751654)(2.3168182,1.848817)(2.4834485,1.4973646)(2.5844367,1.2513481)(2.6854248,0.99026924)(2.796512,0.6337962)(2.872253,0.24719873)(2.9176977,-0.049025342)(2.9530435,-0.5008926)
\pspolygon[linecolor=white, linewidth=0.02, fillstyle=solid,fillcolor=colour1](0.995,-0.7268262)(2.98,-0.7218262)(2.96,-0.36682618)(2.915,0.11817383)(2.825,0.5731738)(2.665,1.1031739)(2.435,1.6431738)(2.23,2.048174)(1.93,2.5031738)(1.625,2.883174)(0.975,2.8881738)
\rput[bl](0.0,3.1481738){\normalsize{time}}
\psframe[linecolor=white, linewidth=0.02, dimen=outer](1.97,-3.3718262)(1.93,-3.4118261)
\psline[linecolor=black, linewidth=0.04, arrowsize=0.05291666666666668cm 2.0,arrowlength=1.4,arrowinset=0.0]{->}(0.98,-0.7218262)(0.98,3.278174)
\rput[bl](3.715,-1.0718262){\normalsize{radius}}
\psline[linecolor=black, linewidth=0.04, arrowsize=0.05291666666666668cm 2.0,arrowlength=1.4,arrowinset=0.0]{->}(0.9850028,-0.724849)(4.604997,-0.71880335)
\pscircle[linecolor=black, linewidth=0.02, fillstyle=solid,fillcolor=black, dimen=outer](1.365,2.7331738){0.07}
\psbezier[linecolor=black, linewidth=0.02](1.375,2.7181737)(1.6309822,2.4986284)(1.9414285,2.2363935)(2.29,1.9131738)
\psbezier[linecolor=black, linewidth=0.02](1.38,2.7181737)(1.2887619,2.6269073)(1.2602092,2.5640993)(1.2,2.4731739)(1.1397908,2.3822484)(1.1218754,2.2112637)(1.125,2.068174)(1.1281246,1.9250841)(1.2159046,1.6845858)(1.45,1.3631738)(1.6840954,1.0417618)(2.625,0.29817382)(2.915,-0.0018261719)
\psbezier[linecolor=black, linewidth=0.02](1.37,2.7181737)(1.3022302,2.5113928)(1.2764322,2.4085267)(1.28,2.2131739)(1.2835679,2.0178208)(1.3922303,1.8563929)(1.6,1.5731739)(1.8077698,1.2899548)(2.598568,0.6928209)(2.865,0.39817384)
\psbezier[linecolor=black, linewidth=0.02](1.38,2.7081738)(1.3993887,2.5865922)(1.4379317,2.4215376)(1.53,2.258174)(1.6220683,2.09481)(1.8043886,1.8815922)(1.95,1.7481738)(2.0956113,1.6147555)(2.2820683,1.4198099)(2.64,1.1881738)
\psbezier[linecolor=black, linewidth=0.02](2.275,1.9431739)(2.9183245,0.9732405)(3.8458443,0.5432405)(4.165,0.28317383)
\psbezier[linecolor=black, linewidth=0.02](2.63,1.2031739)(2.955,0.54817384)(3.93,-0.08182617)(4.17,-0.16182618)
\psbezier[linecolor=black, linewidth=0.02](2.92,0.008173828)(3.245,-0.40682617)(3.43,-0.50682616)(3.795,-0.7168262)
\psbezier[linecolor=black, linewidth=0.02](2.88,0.40317383)(3.235,-0.10682617)(3.7,-0.33682618)(4.185,-0.58182615)
\pspolygon[linecolor=white, linewidth=0.02, fillstyle=solid,fillcolor=colour1](6.59,-0.7218262)(8.575,-0.7168262)(8.555,-0.36182618)(8.51,0.123173825)(8.42,0.5781738)(8.26,1.1081738)(8.03,1.6481738)(7.825,2.0531738)(7.525,2.508174)(7.22,2.8881738)(6.57,2.893174)
\pspolygon[linecolor=white, linewidth=0.02, fillstyle=solid,fillcolor=colour0](8.578291,-0.72180545)(10.98,-0.7218262)(10.98,2.8781738)(7.225,2.883174)(7.4977174,2.5115557)(7.724941,2.1751654)(7.916818,1.848817)(8.083448,1.4973646)(8.184437,1.2513481)(8.285425,0.99026924)(8.396512,0.6337962)(8.472253,0.24719873)(8.517697,-0.049025342)(8.553043,-0.5008926)
\psbezier[linecolor=black, linewidth=0.06](2.98,-0.7318262)(2.9421527,0.3594482)(2.7637274,0.9026519)(2.5186884,1.4611589)(2.2736497,2.019666)(2.0137224,2.4141545)(1.64,2.8581738)
\psbezier[linecolor=black, linewidth=0.06](8.58,-0.7318262)(8.542152,0.3594482)(8.363728,0.9026519)(8.118689,1.4611589)(7.8736496,2.019666)(7.6137223,2.4141545)(7.24,2.8581738)
\psline[linecolor=black, linewidth=0.04, arrowsize=0.05291666666666668cm 2.0,arrowlength=1.4,arrowinset=0.0]{->}(6.5850043,-0.7257062)(11.38,-0.7218262)
\psline[linecolor=black, linewidth=0.04, arrowsize=0.05291666666666668cm 2.0,arrowlength=1.4,arrowinset=0.0]{->}(6.58,-0.7218262)(6.58,3.278174)
\pscircle[linecolor=black, linewidth=0.02, fillstyle=solid,fillcolor=black, dimen=outer](7.51,2.6881738){0.07}
\psbezier[linecolor=black, linewidth=0.02](7.515,2.683174)(8.360982,1.6486284)(9.961429,0.121393524)(10.965,-0.45682618)
\psbezier[linecolor=black, linewidth=0.02](7.52,2.6681738)(8.090982,1.3286284)(9.661428,0.10139353)(10.735,-0.7368262)
\psbezier[linecolor=black, linewidth=0.02](7.491379,2.683174)(7.3714113,2.588552)(7.0882564,2.368435)(6.988886,2.2641666)(6.889515,2.159898)(6.731919,2.0174434)(6.73,1.854093)(6.728081,1.6907426)(6.8701954,1.4936084)(7.135961,1.1694605)(7.4017262,0.84531283)(8.139039,0.18420324)(8.535,-0.18182617)
\psbezier[linecolor=black, linewidth=0.02](8.575,-0.20182617)(8.73,-0.46682617)(8.84,-0.5568262)(9.065,-0.7268262)
\psbezier[linecolor=black, linewidth=0.02](7.51,2.6581738)(7.44223,2.451393)(7.241432,2.1435268)(7.245,1.9481739)(7.248568,1.7528208)(7.55723,1.4413929)(7.735,1.2431738)(7.91277,1.0449548)(8.198568,0.7428209)(8.46,0.47817382)
\psbezier[linecolor=black, linewidth=0.02](8.455,0.47817382)(8.828091,-0.18497618)(9.194091,-0.32712618)(9.775,-0.7368262)
\psbezier[linecolor=black, linewidth=0.02](7.54,2.673174)(7.47723,2.4363928)(7.531432,2.2385268)(7.56,2.1031737)(7.5885677,1.9678209)(7.68723,1.7663928)(7.79,1.6281738)(7.89277,1.4899548)(7.978568,1.4128209)(8.26,1.0881739)
\psbezier[linecolor=black, linewidth=0.02](8.27,1.1131738)(8.678091,0.26002383)(9.59409,-0.32212618)(10.175,-0.7318262)
\rput[bl](6.755,3.1581738){\normalsize{time}}
\rput[bl](10.51,-1.0718262){\normalsize{radius}}
\rput[bl](9.835,3.048174){\normalsize{$(\scrM, g)$}}
\rput[bl](3.0,3.0331738){\normalsize{$(\scrM, g)$}}
\rput[bl](9.375,0.8781738){\normalsize{a}}
\rput[bl](7.675,0.67817384){\normalsize{b}}
\pscircle[linecolor=black, linewidth=0.02, fillstyle=solid,fillcolor=black, dimen=outer](2.27,1.9481739){0.07}
\rput[bl](1.27,2.8681738){\normalsize{$p$}}
\rput[bl](7.725,2.5731738){\normalsize{$p$}}
\rput[bl](2.4,1.8631738){\normalsize{$q$}}
\end{pspicture}
}
\caption{Schematic plot of families of null geodesics ending inside the star (left) and on its outer surface (right).}
\label{figcollapse2}
\end{figure}

We next consider the situation that the null geodesic does not enter the star
(like the geodesic ``a'' on the right of Figure~\ref{figcollapse2}).
Then, again using~\eqref{killconserve} for the Killing field~$K:= \partial_t$, we obtain the conservation law
\[ \la K, \dot{\gamma}(\tau) \ra = e \:, \]
where~$e$ is energy of the geodesic. In the asymptotic end, we can compare this relation
with~\eqref{viszero} to determine the energy to~$e=1/\varepsilon$. Hence
\[ \dot{\gamma}_0(\tau) = \la K, \dot{\gamma}(\tau) \ra = \frac{1}{\varepsilon} \:. \]
In order to clarify what this means, it is useful to choose a local reference frame
at~$p=\gamma(\tau)$. Then the time component in this local reference frame is given
by~$\la T, \dot{\gamma} \ra$, where~$T$ is a unit vector proportional to~$\partial_t$. A direct computation gives
\beq \label{locframe}
\la T, \dot{\gamma}(\tau) \ra = \left( 1 - \frac{2M}{r(\tau)} \right)^{-\frac{1}{2}} 
\dot{\gamma}_0(\tau) = \left( 1 - \frac{2M}{r(\tau)} \right)^{-\frac{1}{2}} \: \frac{1}{\varepsilon}\:.
\eeq
When approaching the star, this inner product becomes larger, implying that the resulting
length scale~$\varepsilon_p$ (as defined in a local reference frame by~\eqref{epsprel}) becomes smaller.
As a consequence, the gravitational coupling also becomes smaller.
The radial dependence in~\eqref{locframe} simply is the usual {\em{red shift effect}}.
We thus conclude that the red shift effect {\em{makes the DGC function smaller}}.

The situation becomes more interesting if the null geodesic crosses the surface of the star.
We first consider the situation that the point~$p$ lies in the interior of the star
(see the left plot in Figure~\ref{figcollapse2}). For computational purposes, it is most convenient
to parametrize the null geodesic such that
\[ \tilde{\gamma}(0)=p \qquad \text{and} \qquad \dot{\tilde{\gamma}}^0(0) = 1 \:. \]
Following the geodesic in Minkowski space backwards in time, it crosses the shell of matter
at a point~$q \in \scrM$
(note that the fact that the curves in Figure~\ref{figcollapse2} ``turn around'' near~$r=0$
is a consequence of the usual centrifugal potential).
Transforming to a chart where the metric is continuous, the tangent vector~$\dot{\tilde{\gamma}}$ inside is
uniquely matched to a corresponding tangent vector~$\dot{\gamma}$ outside.
Taking~$q$ and~$\dot{\gamma}$ as initial data, one can solve the geodesic equation
in Schwarzschild to obtain~$\gamma(\tau)$ for large negative~$\tau$.
By reparametrizing, one can then satisfy the condition~\eqref{viszero}.

\begin{figure}
\includegraphics[width=8cm]{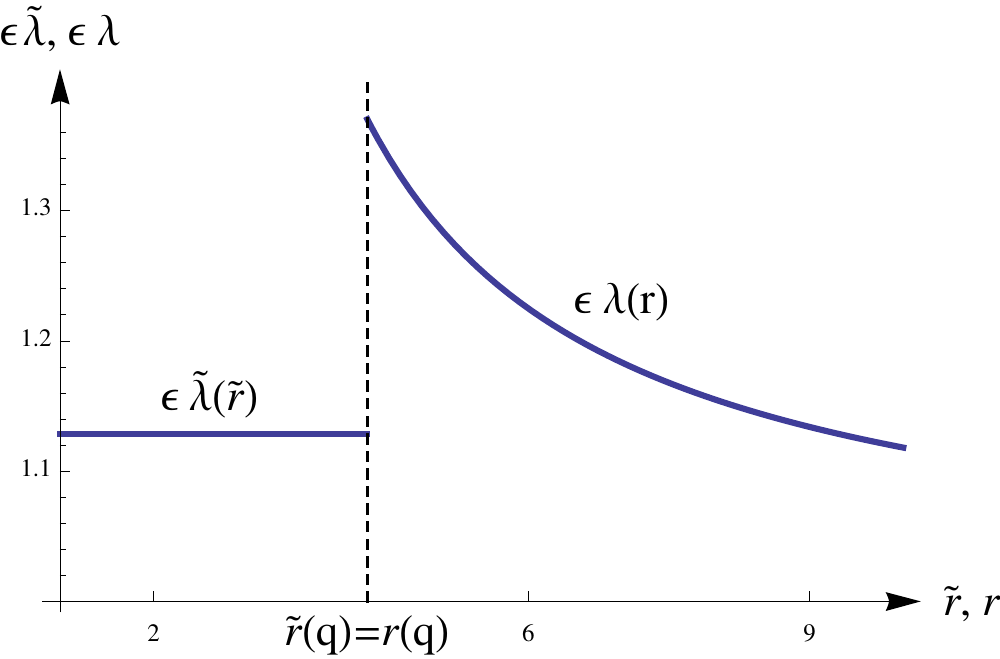}
\caption{Refraction of a null geodesic on the surface of the star.}
\label{figrefract}
\end{figure}

The matching of the tangent vector can be understood as describing a refraction
of the geodesic on the surface of the star. In order to explain the effect of this refraction,
we now give a numerical example\footnote{All our numerics was carried
out using the ODE solver in {\textsf{Mathematica}}. The {\textsf{Mathematica}} files are included as ancillary
files to the arXiv submission of this paper.}.
We consider a star of mass~$M=1$ and for simplicity a geodesic without angular momentum
which moves radially outwards (the geodesics with angular momentum have the same qualitative
behavior). In our example, the geodesic hits the surface of the star at a radius~$r=\tilde{r} \approx 4.28$, where the
velocity of the surface of the star is about 0.56 times the speed of light. In Figure~\ref{figrefract},
the quantities
\[ \tilde{\lambda} := \dot{\tilde{\gamma}}^0 \qquad \text{and} \qquad \lambda := \la T, \dot{\gamma} \ra \]
are shown as functions of the radial variables~$\tilde{r}$ and~$r$, respectively,
where~$T$ is again the future-directed unit vector proportional to~$\partial_t$.
The behavior of~$\lambda$ outside the star again corresponds to the red shift effect~\eqref{locframe}.
One sees that the {\em{refraction partly compensates the red shift effect}}, because the
function~$\lambda$ is again smaller inside the star.
A careful analysis of stars whose surface speed is very close to the speed of light shows that
this effect occurs even in the ultrarelativistic scenario. As a consequence, inside the star,
the DGC function stays bounded away from zero even if the red shift effect on the 
surface of the star becomes arbitrarily large.
However, it does not seem possible to arrange that inside the star~$\lambda$ is smaller than~$1/\varepsilon$.
This implies that inside the star, the DGC function~$\kappa_\vol$ is necessarily smaller than at infinity.

\begin{figure}
\psscalebox{1.0 1.0} 
{
\begin{pspicture}(0,-3.1427205)(11.395,3.1427205)
\definecolor{colour0}{rgb}{0.8,0.8,0.8}
\psframe[linecolor=white, linewidth=0.02, dimen=outer](2.99,-3.1027203)(2.95,-3.1427205)
\rput[bl](1.615,1.0972797){\normalsize{$p$}}
\pscircle[linecolor=black, linewidth=0.06, fillstyle=solid,fillcolor=colour0, dimen=outer](0.8,1.5472796){0.8}
\pscircle[linecolor=black, linewidth=0.02, fillstyle=solid,fillcolor=black, dimen=outer](1.71,1.5522796){0.07}
\psbezier[linecolor=black, linewidth=0.04](1.696215,1.5355495)(3.244821,1.7376336)(4.3974857,2.108805)(6.158785,3.1240098)
\psbezier[linecolor=black, linewidth=0.04](1.725,1.5572796)(3.16,1.9822797)(4.4,2.5472796)(5.105,3.1272798)
\psline[linecolor=black, linewidth=0.02](1.72,1.5272796)(4.005,1.5472796)
\psline[linecolor=black, linewidth=0.02](5.365,2.6872797)(6.29,2.6872797)
\psarc[linecolor=black, linewidth=0.02, dimen=outer](2.2575,1.5397797){1.1525}{0.0}{17.0}
\rput[bl](3.5,1.6222796){\normalsize{$\alpha$}}
\psarc[linecolor=black, linewidth=0.02, dimen=outer](4.7775,2.6897798){1.3325}{0.0}{17.0}
\rput[bl](6.295,2.8022797){\normalsize{$\phi_\text{in}$}}
\end{pspicture}
}
\caption{Schematic plot of null geodesics in the equatorial plane.}
\label{figangles}
\end{figure}
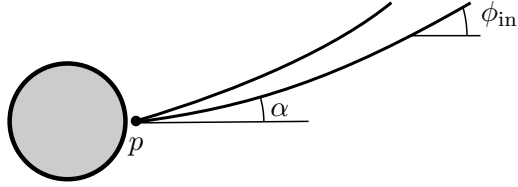

\begin{figure}
\begin{center}
\includegraphics[width=6.5cm]{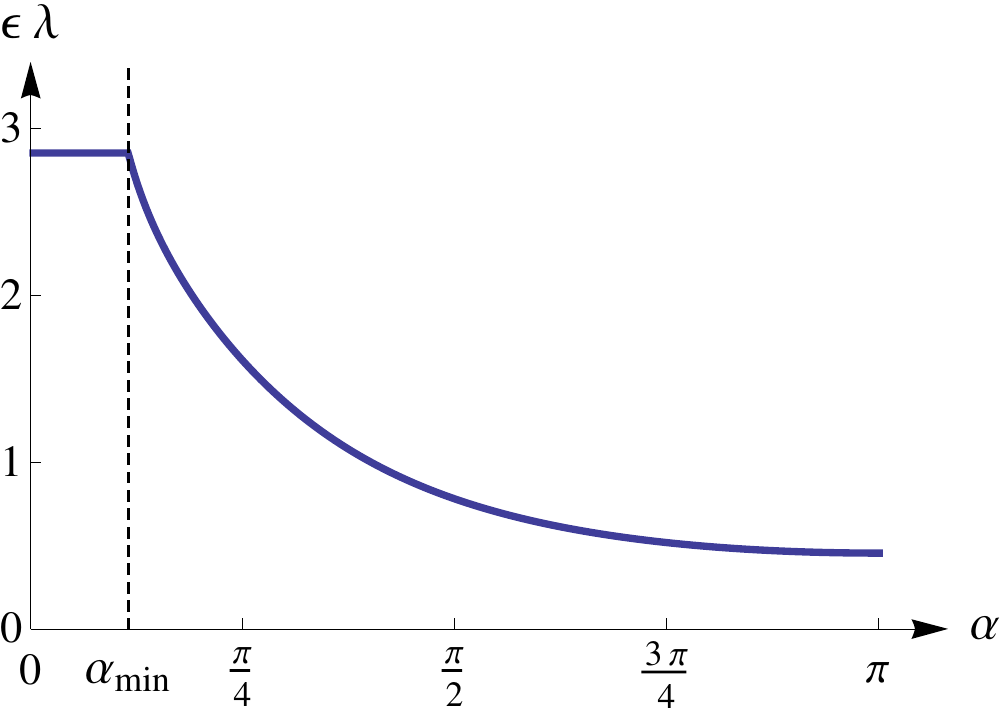} $\quad$
\includegraphics[width=6.5cm]{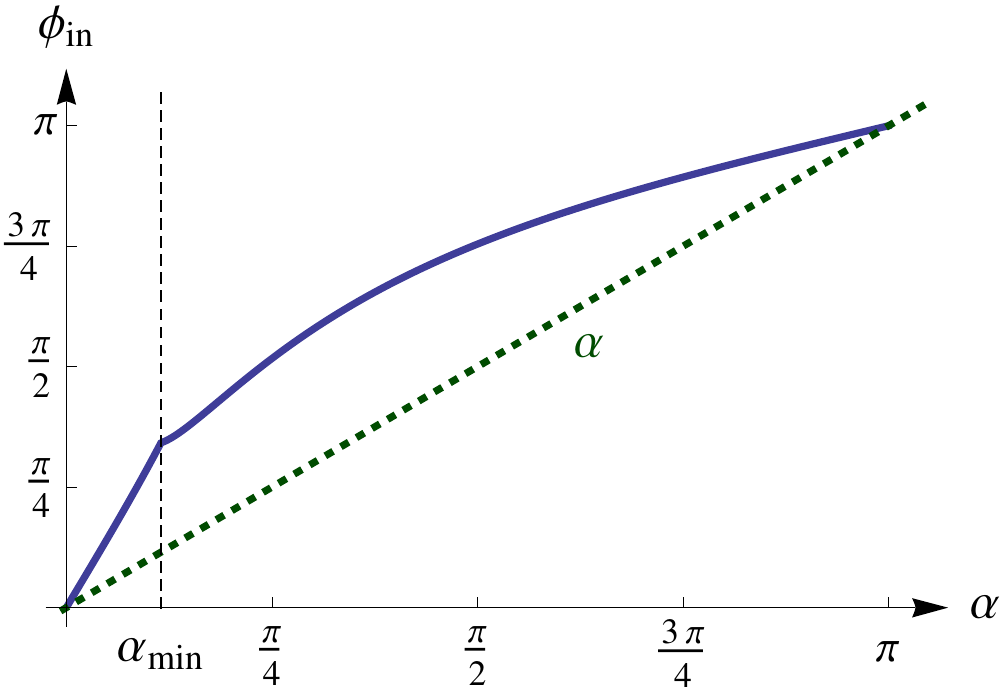}
\end{center}
\caption{The functions $\lambda(\alpha)$ and~$\phi_\text{in}(\alpha)$ on the outer surface of the star.}
\label{figalpha}
\end{figure}

We next consider the situation that~$p$ is on the outer surface of the star (see the right plot
in Figure~\ref{figcollapse2}). Now we must distinguish the geodesics which do not meet the star
(like the geodesic ``a'' on the right of Figure~\ref{figcollapse2}) from those which enter and leave
the star (like the geodesic ``b'' on the right of Figure~\ref{figcollapse2}).
To this end, we denote the angle between the geodesic and the line~$\{\phi=0\}$
in the equatorial plane by~$\alpha \in [0, \pi]$ (see Figure~\ref{figangles}).
Moreover, we let~$\phi_\text{in} = \lim_{\rightarrow -\infty} \phi(\tau)$ be the angle in the asymptotic end.
Then there is an angle~$\alpha=\alpha_{\min}$ for which the null geodesic just touches the surface of the star.
On the left of Figure~\ref{figalpha},
the function~$\lambda$ at the point~$p$ is plotted as a function of~$\alpha$,
for the same star as considered in Figure~\ref{figrefract} (the speed of the surface of the star at~$p$
is about $0.7$ times the speed of light).
Likewise, on the right of Figure~\ref{figalpha}, the angle~$\phi_\text{in}$ is shown.
Let us discuss these plots. If~$\alpha<\alpha_{\min}$, the geodesic does not meet the star.
Therefore, $\lambda$ is described purely by the red shift effect described in~\eqref{locframe}.
The reason why~$\phi_\text{in} > \alpha$ (see the green dotted line on the right of Figure~\ref{figalpha})
is that the geodesics are bent by the gravitational attraction
(as indicated in Figure~\ref{figangles}).
If~$\alpha > \alpha_{\min}$, the geodesic is refracted twice, once when it enters the star
and again when it leaves. The effect of this
{\em{twofold refraction}} is that at the space-time point~$p$, the function~$\lambda$
can indeed become smaller than~$1/\varepsilon$.
Thus the double reflection can overcompensate the red shift effect.

We now explain the effect on the DGC function at the point~$p$.
The volume~$\mu_p(D_p\scrL)$ is computed by (for details see Appendix~\ref{appcone})
\beq \label{mup}
\mu_p(D_p\scrL) = 2 \int_{S^2} \lambda^2\: d\mu_{S^2} = 4 \pi \int_{-1}^1 \lambda(\alpha)^2\: d\cos \alpha \:.
\eeq
In the region~$\alpha < \alpha_{\min}$, the integrand is large, giving a large contribution to the
integral provided that~$\alpha_{\min}$ is not too small.
This suggests that $\kappa_\vol$ becomes small (see~\eqref{kappavol}),
in agreement with the above discussion of the red shift effect.
In the region~$\alpha>\alpha_{\min}$, on the other hand, the integrand becomes small, thus giving a
small contribution to the integral. In the example shown in Figure~\ref{figalpha},
we find numerically that~$\kappa_\vol = 0.72 \times \varepsilon^2/(8 \pi)$.
It turns out that considering ultrarelativistic situations where the speed of surface of the star is very close to the
speed of light, the integrand in~\eqref{mup} becomes very large near~$\alpha=0$
(again due to the red shift effect). But~$\alpha_{\min}$ becomes small in such a way that the integral~\eqref{mup}
remains bounded. We conclude that, although the double reflection can overcompensate the red shift
effect, it does not seem possible to arrange that~$\kappa_\vol(p)$ becomes larger than at infinity.

\begin{figure}
\begin{center}
\includegraphics[width=8cm]{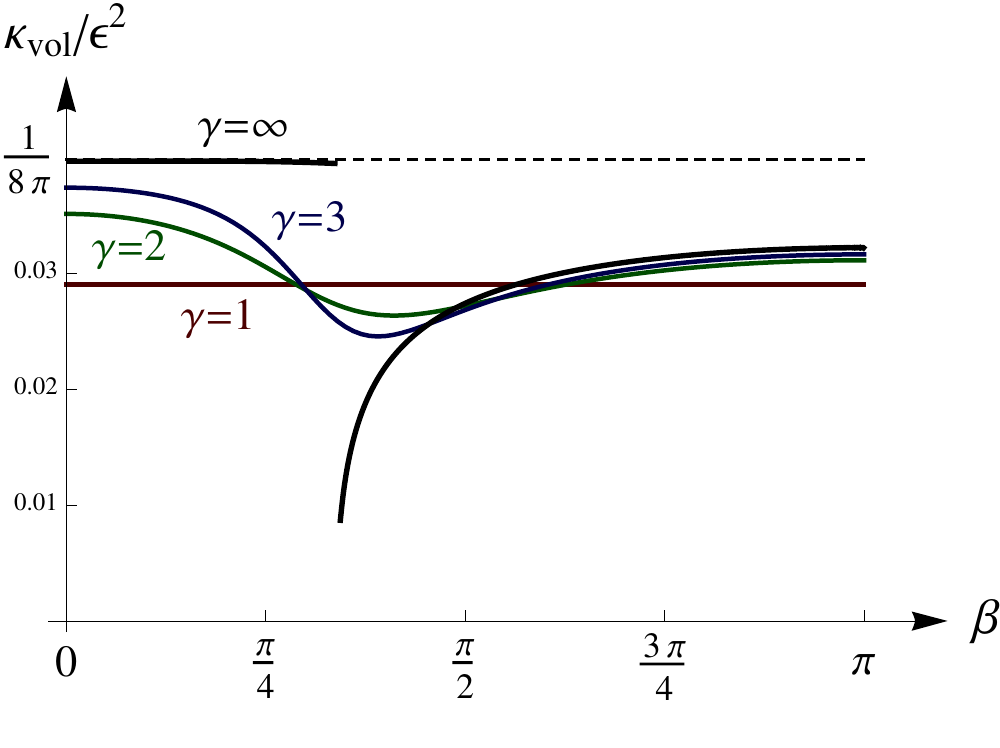}
\end{center}
\caption{The dynamical coupling~$\kappa_\vol$ as a function of the parameters in~\eqref{par}.}
\label{figkappa1}
\end{figure}

We next discuss the effect if the star moves relative to the reference frame
distinguished by the regularization. Then the null geodesics should be parametrized
in the asymptotic end according to~\eqref{LMinkv}. Therefore, the condition~\eqref{viszero}
for the future-directed null geodesics must be replaced by
\beq \label{vnonzero}
\la v, \dot{\gamma}(-\infty) \ra = \frac{1}{\varepsilon} \:,
\eeq
where~$\dot{\gamma}(-\infty) := \lim_{\tau \rightarrow -\infty} \dot{\gamma}(\tau)$
is a vector in Minkowski space, and $v$ is again the unit vector~\eqref{vdef}.
Since the transformation from~\eqref{viszero} to~\eqref{vnonzero} is described by
a linear reparametrization, the volume~$\mu_p(D_p\scrL)$ is obtained
by inserting an additional factor into the integral in~\eqref{mup},
\[ \mu_p(D_p\scrL) = 2 \int_{S^2} \frac{\lambda^2}{\varepsilon^2 \,\la v, \dot{\gamma}(-\infty)\ra^2}\: d\mu_{S^2} \:. \]
This additional factor has the effect that the main contribution to the integral
comes when the inner product~$\la v, \dot{\gamma}(-\infty)\ra$ is small,
which is the case if the vectors~$\vec{v}$ and~$\dot{\vec{\gamma}}(-\infty)$
are pointing in the same spatial direction.
As a consequence, the effective gravitational coupling depends on the parameters
\beq \label{par}
\beta := \sphericalangle \big(\vec{v}, -\dot{\vec{\gamma}}(-\infty) \big) \in [0, \pi]
\qquad  \text{and} \qquad \gamma := v^0 \:.
\eeq
When computing the resulting integral over the sphere, by a rotation one can
clearly arrange that the vector~$\vec{v}$ lies in the equatorial plane.
But then the vector~$\dot{\vec{\gamma}}(-\infty)$ can no longer be arranged to
also lie in the equatorial plane. However, the resulting integral over the angle to the equatorial plane
can be computed explicitly to obtain an expression of the form
\beq \label{mupW}
\mu_p(D_p\scrL) = 8 \pi \int_{-1}^1 \lambda(\alpha)^2\: W\big(\phi_\text{in}(\alpha), \beta, \gamma \big)
\:d\cos \alpha \:,
\eeq
where
\[ W(\phi_\text{in}, \beta, \gamma)
= \frac{1}{2}\frac{\gamma - \sqrt{\gamma^2-1}\, \cos \beta \,\cos \phi_\text{in}}
{\left(\big(\gamma -\sqrt{\gamma^2-1}\, \cos \beta \,\cos \phi_\text{in} \big)^2- 
\big(\gamma^2-1 \big)\, \sin^2 \beta\, \sin^2 \phi_\text{in} \right)^{3/2}} \:. \]
In Figure~\ref{figkappa1},
the dynamical coupling~$\kappa_\vol$ is shown as a function of~$\beta$ and~$\gamma$,
again for the same star as in Figure~\ref{figrefract}. Thus~$\kappa_\vol$ is largest if~$\gamma$ is large,
and if the angle~$\beta$ is small.

The limiting case~$\gamma=\infty$ shown in Figure~\ref{figkappa1} is indeed a major
technical simplification. The reason is that
\beq \label{Wlimit}
\lim_{\gamma \rightarrow \infty} W(\phi_\text{in}, \beta, \gamma) = \frac{1}{\sin \phi_\text{in}}
\: \delta( \phi_\text{in} - \beta)
\eeq
(with convergence as a distribution). Even without going through the detailed computations,
this result can be understood immediately as follows:
In the limiting case, one only gets a contribution if~$\vec{v}$ and~$\dot{\vec{\gamma}}(-\infty)$
are parallel, implying that~$\beta = \phi_\text{in}$. This gives rise to the $\delta$-distribution in~\eqref{Wlimit}.
Moreover, in Minkowski space,
we know that~$\phi_\text{in}(\alpha)=\alpha$ (see Figure~\ref{figangles}), and
the invariance of~$\mu_p(D_p\scrL)$ under Lorentz boosts implies that
\[ \int_{-1}^1 W\big(\alpha, \beta, \gamma \big)
\:d\cos \alpha = 1 \qquad \text{for all~$\beta, \gamma$}\:, \]
explaining the factor~$1/\sin \phi_\text{in}$ in~\eqref{Wlimit}.
Using~\eqref{Wlimit} in~\eqref{mupW}, we obtain
\[ \lim_{\gamma \rightarrow \infty} \mu_p(D_p\scrL)
= 8 \pi \int_{-1}^1 \lambda(\alpha)^2
\: \delta \big( \phi_\text{in}(\alpha) - \beta \big) \: \frac{\sin \alpha}{\sin \phi_\text{in}} \:d\alpha
= 8 \pi  \: \frac{\lambda(\alpha)^2}{\phi_\text{in}'(\alpha)}\;
\frac{\sin \alpha}{\sin \beta} \:, \]
where the last expression is to be evaluated at~$\alpha$ with~$\phi_\text{in}(\alpha) = \beta$.
Using~\eqref{kappavol}, we obtain
\beq \label{gammainf}
\kappa_\vol = \frac{1}{8 \pi}  \: \frac{\phi_\text{in}'(\alpha)}{\lambda(\alpha)^2}\;
\frac{\sin \beta}{\sin \alpha} \bigg|_{\alpha \text{ with } \phi_\text{in}(\alpha) = \beta} \:.
\eeq
This function is much easier to compute numerically than the integral~\eqref{mupW}.
It gives the plot for~$\gamma=\infty$ in Figure~\ref{figkappa1}.
We note that this function has a discontinuity at the angle~$\beta = \phi_\text{in}(\alpha_{\min})$
(where~$\alpha_{\min}$ is the angle at which the null geodesic is tangential to the
surface of the star). This comes about because the function~$\phi_\text{in}(\alpha)$ is
continuous but not differentiable at~$\alpha_{\min}$ (see the right plot in Figure~\ref{figalpha}),
so that its derivative in~\eqref{gammainf} is not continuous.

The appearance of the function~$\phi_\text{in}'(\alpha)$ in~\eqref{gammainf} teaches the general
lesson that the DGC function also depends on the {\em{scattering}} of the
null geodesics by the gravitational field (as shown in Figure~\ref{figangles}).
Thus the two relevant effects for understanding the collapse of a star are
refraction and scattering of the null geodesics. Thinking of a realistic star, these effects
do not need to originate from the star itself, but could also be caused by stars or matter
in its surrounding. Analyzing these effect further goes beyond the scope of this paper.

\begin{figure}
\begin{center}
\includegraphics[width=6.5cm]{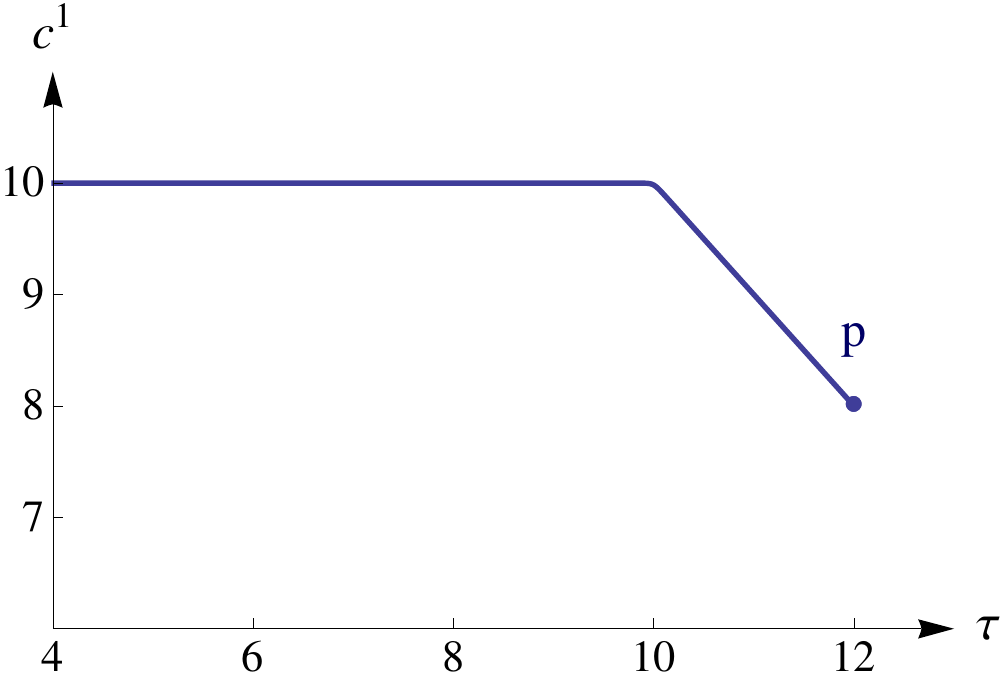} $\quad$
\includegraphics[width=7cm]{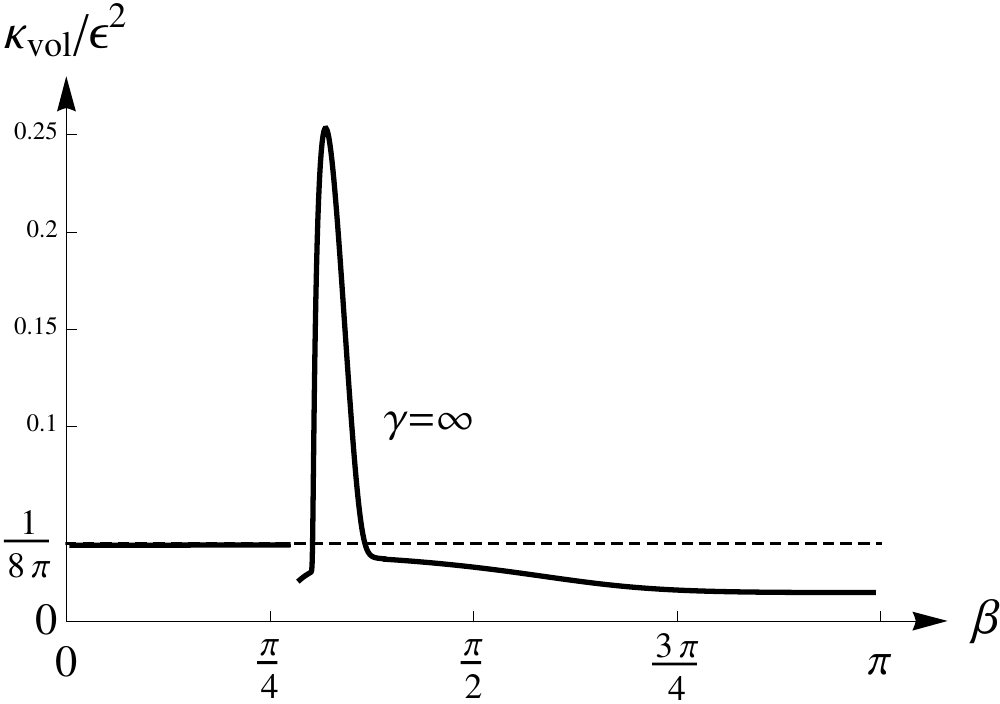}
\end{center}
\caption{The dynamical coupling~$\kappa_\vol$ in an example involving non-gravitational forces.}
\label{figkappa2}
\end{figure}

We analyzed the DGC function~\eqref{gammainf} numerically for different shells of matter.
In order to model non-gravitational forces acting on the star (like pressure, electromagnetic forces,
etc.), we also considered the
situation that the curve~$c(\tau)$ in~\eqref{ctau} describing the outer surface of
the star is no longer a geodesic. An interesting point of our findings is that
in generic situations, the DGC function~$\kappa_\vol$ becomes larger than at infinity.
Indeed, it can even be arranged to become arbitrarily large.
This is illustrated in Figure~\ref{figkappa2} for a star of mass~$M=2$ which is first static and then
collapses with constant radial component of the velocity vector~$\dot{c}^1(\tau)$,
where the velocity of the surface at~$p$ is~$0.7$ times the speed of light.
This means that DGC can lead to the effect that the gravitational field generated by a star
is stronger than in Einstein's theory. Therefore, looking at its gravitational field, the star
appears to be heavier. This effect might give an explanation for {\em{dark matter}}.
But clearly, a systematic study of the gravitational coupling for
realistic collapsing stars goes beyond the scope of this paper.

In the previous analysis, for simplicity we restricted attention to the DGC function~$\kappa_\vol$.
We finally remark how the {\em{results carry over to the function~$\kappa(p)$}} in~\eqref{kappadyn}:
First of all, the functions~$\kappa(p)$ and~$\kappa_\vol(p)$ coincide for linearized gravity,
so that the differences concern only the nonlinear effects in strong gravitational fields.
Qualitatively speaking, the functions~$\kappa(p)$ and~$\kappa_\vol(p)$ are approximately equal
as long as the Gaussian curvature~$K(p)$ is nearly constant.
In particular, the results shown in Figure~\ref{figkappa1} remain valid for~$\kappa(p)$, up to
small corrections. Likewise, the results shown in Figure~\ref{figkappa2} remain qualitatively the same for~$\kappa(p)$,
because the peak of~$\kappa_\vol$ is not caused by a fluctuation of~$\lambda$, but 
by a ``spike'' of the factor~$\phi_\text{in}'(\alpha)$ in~\eqref{gammainf}.
But clearly, the numerical values for~$\kappa(p)$ are a little bit smaller than those for~$\kappa_\vol(p)$
(see Lemma~\ref{lemmaholder}).

\section{Dynamical Gravitational Coupling in Weak Fields} \label{secweak}
\subsection{Linearized Gravity with Dynamical Gravitational Coupling} \label{seclin}
We now consider linearized gravity. Thus we write the metric as
\[ g_{jk} = \eta_{jk} + h_{jk} \]
and take into account the deviation from the Minkowski metric~$h_{jk}$ only linearly.
Let~$\gamma(\tau)$ be a parametrized null geodesic with~$\gamma(0) = x$.
In Minkowski space, this null geodesic can be written as
\[ \gamma(\tau) = x + \tau u \qquad \text{with} \qquad \eta_{jk} u^j u^k = 0 \:. \]
Let us compute how this null geodesic is modified by the gravitational field.
First, in order for the tangent vector~$\dot{\gamma}(0)$ to be a null vector, we choose
the initial values
\[ \gamma(0) = x \qquad \text{and} \qquad \dot{\gamma}^j(0) = u^j - \frac{1}{2}\: h^{jk}(x) \, u_k \]
(where we lower and raise indices with respect to the Minkowski metric).
Namely, with this choice,
\[ g_{jk}(x)\: \dot{\gamma}^j(0) \:\dot{\gamma}^k(0)
= \eta_{jk}\,u^j u^k + h_{jk}(x)\:u^j u^k  + 2 \eta_{jk} \,u^j\, \Big( - \frac{1}{2}\: h^{kl}(x) \, u_l \Big) = 0 \:. \]
The geodesic equation becomes
\[ \ddot{\gamma}^j(\tau) = -\Gamma^j_{kl}\big(\gamma(\tau)\big)\: \dot{\gamma}^k(\tau)\: \dot{\gamma}^l(\tau)\:. \]
Since the Christoffel symbols vanish in Minkowski space, the perturbation of the 
geodesic is described as follows,
\begin{align}
\gamma(\tau) &= x + \tau u + \delta \gamma(\tau) \label{gpert} \\
\delta \gamma(0) &= 0 \:,\qquad
\delta \dot{\gamma}^j(0) = -\frac{1}{2}\: h^{jk}(x) \, u_k \\
\delta \ddot{\gamma}^j(\tau) &= -\Gamma^j_{kl}\big(x + \tau u \big)\: u^k u^l \:.
\end{align}
This differential equation can be solved immediately by integration,
\begin{align}
\delta \dot{\gamma}^j(\tau) &= -\frac{1}{2}\: h^{jk}(x) \, u_k
- \int_0^\tau \Gamma^j_{kl}\big(x + \alpha u \big)\: u^k u^l \: d\alpha \label{dgdot} \\
\delta \gamma^j(\tau) &= -\frac{\tau}{2}\: h^{jk}(x) \, u_k
- \int_0^\tau d\beta \int_0^\beta d\alpha\: \Gamma^j_{kl}\big(x + \alpha u \big)\: u^k u^l \notag \\
&=-\frac{\tau}{2}\: h^{jk}(x) \, u_k
- \int_0^\tau (\tau-\alpha)\: \Gamma^j_{kl}\big(x + \alpha u \big)\: u^k u^l \: d\alpha \:, \label{dg}
\end{align}
where in the last step we transformed the double integral by
\begin{align*}
\int_0^\tau d\beta \int_0^\beta f(\alpha) \:d\alpha
&= \int_0^\tau d\beta \int_0^\tau d\alpha\: \Theta(\beta - \alpha)\: f(\alpha) \\
&= \int_0^\tau d\alpha \: f(\alpha)\int_0^\tau d\beta\: \Theta(\beta - \alpha) 
= \int_0^\tau (\tau-\alpha)\: f(\alpha)\:d\alpha \:.
\end{align*}

Combining~\eqref{gpert} with~\eqref{dgdot} and using the explicit form of the Christoffel symbols
\[ \Gamma^j_{kl} = \frac{1}{2}\: \eta^{jp} \left( \partial_k h_{lp} + \partial_l h_{kp} - \partial_p h_{kl} \right) \:, \]
we conclude that
\begin{align*}
\dot{\gamma}^j(\tau) &= u^j -\frac{1}{2}\: h^{jk}(x) \, u_k
- \int_0^\tau \Gamma^j_{kl}\big(x + \alpha u \big)\: u^k u^l \: d\alpha \\
&= u^j -\frac{1}{2}\: h^{jk}(x) \, u_k
- \int_0^\tau \: \big(u^k \partial_k h_l^{\:\:j}\big)(x+\alpha u)\: u^l \: d\alpha
+\frac{1}{2} \int_0^\tau \partial^j h_{kl}(x+\alpha u) \: u^k u^l \: d\alpha \\
&= u^j -\frac{1}{2}\: h^{jk}(x) \, u_k
- \int_0^\tau \: \frac{d}{d\alpha} h^{jk}(x+\alpha u)\: u_k \: d\alpha
+\frac{1}{2} \int_0^\tau \partial^j h_{kl}(x+\alpha u) \: u^k u^l \: d\alpha \\
&= u^j +\frac{1}{2}\: h^{jk}(x) \, u_k - h^{jk}(x+\tau u) \, u_k
+\frac{1}{2} \int_0^\tau \partial^j h_{kl}(x+\alpha u) \: u^k u^l \: d\alpha \:.
\end{align*}
It is instructive to verify the formula as follows:
\begin{itemize}[leftmargin=1.5em]
\itemD $\dot{\gamma}(\tau)$ is a null vector:
\begin{align*}
g&_{ij}\big(\gamma(\tau) \big) \: \dot{\gamma}^i(\tau)\: \dot{\gamma}^j(\tau)
= h_{ij}(x+\tau u)\: u^i u^j + 2 u_j \: \delta \dot{\gamma}^j(\tau) \\
&= h_{ij}(x+\tau u)\: u^i u^j  
+ u_j\, h^{jk}(x) \, u_k -2 u_j\:h^{jk}(x+\tau u) \, u_k \\
&\qquad +\int_0^\tau u_j \partial^j h_{kl}(x+\alpha u) \: u^k u^l \: d\alpha \\
&= -h_{kl}(x+\tau u)\: u^k u^l + h_{kl}(x)\: u^k u^l
+\int_0^\tau \frac{d}{d\alpha} h_{kl}(x+\alpha u) \: u^k u^l \: d\alpha = 0 \:.
\end{align*}
\itemD For a conformal transformation, we obtain
\begin{align}
h_{jk} &= h(x)\: \eta_{jk} \label{linconform} \\
\dot{\gamma}(\tau) &= u +\frac{1}{2}\: h(x)\, u - h(x+\tau u)\: u
= \Big(1 - h(x+\tau u) \Big) u + \frac{1}{2}\: h(x)\, u\:,
\end{align}
in agreement with the computations in Section~\ref{secconftrans}.
\end{itemize}

We now consider the situation when the point~$x$ is fixed
and~$\tau \rightarrow -\infty$, so that the point~$y:=x+\tau u$ 
moves along the null geodesic to null infinity in the past.
Furthermore, we assume that the gravitational field vanishes in the distant past.
We then obtain
\begin{align}
\dot{\gamma}^j(-\infty) &= u^j +\frac{1}{2}\: h^{jk}(x) \, u_k
-\frac{1}{2} \int_{-\infty}^0 \partial^j h_{kl}(x+\alpha u) \: u^k u^l \: d\alpha \notag \\
&= u^j +\frac{1}{2} \int_{-\infty}^0 \big( \partial_l h^j_{\;\:k}
- \partial^j h_{kl} \big)(x+\alpha u) \: u^k u^l \: d\alpha \:. \label{dg2}
\end{align}
We now let~$\scrL$ be the set of all null geodesics which at infinity goes over to
the family~\eqref{LMinkv} considered in Minkowski space.
We now compute the resulting set~$D_x \scrL$. It is most convenient to
work in a reference frame where~$g_{ij}(x) = \eta_{ij}(x)$. 
In this reference frame, we parametrize the set~$D_x \scrL$ as
\[ D_x \scrL = \left\{ \pm \lambda(\vec{n}) \begin{pmatrix} 1 \\ \vec{n} \end{pmatrix} 
\quad \text{with} \quad
\big|\vec{n} \big| = 1  \right\} \:, \]
where~$\lambda : S^2 \rightarrow \R^+$ is a non-negative function on the unit sphere.
In order to determine~$\lambda$, for given~$\vec{n}$ we consider the null geodesic~\eqref{dg2}
with~$u=(1,\vec{n})$. Reparametrizing the null geodesic such as to satisfy~\eqref{LMinkv}, we obtain
\[ \lambda = \frac{1}{\varepsilon\:\la v, \dot{\gamma}(-\infty)\ra} 
= \frac{1}{\varepsilon\: \la v, u \ra} - \frac{1}{2 \varepsilon \:\la v, u \ra^2}
\int_{-\infty}^0 \big( \partial_l h_{jk} - \partial_j h_{kl} \big)(x+\alpha u) \: v^j\, u^k u^l \: d\alpha
\:. \]
The volume~$\mu_x(D_x \scrL)$ is computed as follows (for details see Appendix~\ref{appcone}):
\begin{align*}
&\mu_x(D_x \scrL) = 2 \int_{S^2} \lambda^2\: d\mu_{S^2}(\vec{n}) \\
\;\;\;&= \frac{2}{\varepsilon^2} \int_{S^2} \frac{d\mu_{S^2}(\vec{n})}{\la v, u \ra^2} 
- \frac{2}{\varepsilon^2} \int_{S^2} \frac{d\mu_{S^2}(\vec{n})}{\la v, u \ra^3}
\int_{-\infty}^0 \big( \partial_l h_{jk} - \partial_j h_{kl} \big)(x+\alpha u) \: v^j u^k u^l \: d\alpha \\
&= \frac{8 \pi}{\varepsilon^2}
- \frac{2}{\varepsilon^2} \int_{S^2} \frac{d\mu_{S^2}(\vec{n})}{\la v, u \ra^3}
\int_{-\infty}^0 \big( \partial_l h_{jk} - \partial_j h_{kl} \big)(x+\alpha u) \: v^j u^k u^l \: d\alpha \:.
\end{align*}
Using~\eqref{kappavol} together with the fact that~$\kappa$ and~$\kappa_\vol$ coincide
for linearized gravity, we obtain the dynamical gravitational coupling
\beq \label{kappafinal}
\kappa = \kappa_\vol = \frac{\varepsilon^2}{8 \pi} \left( 1 + \frac{1}{4 \pi}
\int_{S^2} \frac{d\mu_{S^2}(\vec{n})}{\la v, u \ra^3}
\int_{-\infty}^0 \big( \partial_l h_{jk} - \partial_j h_{kl} \big)(x+\alpha u) \: v^j \,u^k u^l \: d\alpha \right) \:.
\eeq
We remark that for a conformal transformation~\eqref{linconform}, the above formula simplifies to
\begin{align*}
\kappa = \kappa_\vol &= \kappa_0 \left(1+
\frac{1}{4 \pi} \int_{S^2} \frac{d\mu_{S^2}(\vec{n})}{\la v, u \ra^3} \int_{-\infty}^0 
d\alpha\, \Big( u^l \partial_l h(x+\alpha u) \: \la v, u \ra \Big) \right) \\
&= \kappa_0 \left(1+
\frac{1}{4 \pi} \int_{S^2} \frac{d\mu_{S^2}(\vec{n})}{\la v, u \ra^2} \:
h(x) \right) = \kappa_0 \:\big(1+ h(x) \big) \:,
\end{align*}
in agreement with the computations in Section~\ref{secconftrans}.

Now the linearized Einstein equations with DGC
can be formulated similar to the presentation~\cite[Section~5.1]{straumann2}.
Choosing the Hilbert gauge
\[ \partial_i \gamma^{ij} = 0 \qquad \text{where} \qquad
\gamma_{ij} := h_{ij} - \frac{1}{2}\: h\, \eta_{ij} \:, \]
the linearized Einstein equations without DGC are
\[ \Box \gamma_{jk} = -16 \pi G\, T_{jk} \:. \]
DGC is included by the replacement~$8 \pi G \rightarrow \kappa(x)$.
Thus for the linearized Einstein equations with DGC we obtain
\beq \label{linh}
\Box h_{jk}(x) = -2 \kappa(x)\,\Big( T_{jk}(x) -\frac{1}{2}\: T\: \eta_{jk}(x) \Big) \:.
\eeq

\subsection{The Newtonian Limit} \label{secnewton}
In order to describe the Newtonian limit, we assume that the only relevant component of
the energy-momentum tensor is the $00$ component, i.e.
\begin{align*}
T_{jk} &= \text{diag} \big( \rho, 0,0,0 \big) \\
T_{jk} -\frac{1}{2} \:T\: g_{jk} &= \frac{\rho}{2} \:\text{diag} \big( 1, 1,1,1 \big) \:.
\end{align*}
Moreover, we assume that the time derivatives in~\eqref{linh} can be neglected, so that
\[ -\Delta h_{jk} = -16 \pi G\, \left( T_{jk} -\frac{1}{2}\: T\: \eta_{jk} \right) . \]
This gives the relations
\[ h_{jk} = 2 \phi\, \text{diag} \big( 1, 1,1,1 \big) 
= 2 \phi\, \big( 2 \delta_{i0} \delta_{j0} - \eta_{jk} \big)  \:, \]
where~$\phi$ is the Newtonian potential,
\[ \phi(t,\vec{x}) = - G \int_{\R^3} \frac{\rho(t,\vec{x})}{|\vec{x}-\vec{y}|} \: d^3y \:. \]
Note that we follow the usual convention that the Newtonian potential is negative, because
it describes an attractive force.

Using the above formulas in~\eqref{kappafinal}, we obtain
\begin{align*}
\frac{\kappa(x)}{\kappa_0} = \frac{\kappa_\vol(x)}{\kappa_0} = 1 + \frac{1}{2 \pi}
\int_{S^2} \frac{d\mu_{S^2}(\vec{n})}{\la v, u \ra^3}
\big( 2 u^j \,v^0 - u^j\, \la v,u \ra - 2 v^j \big) 
\int_{-\infty}^0 \partial_j \phi(x+\alpha u) \: d\alpha \:,
\end{align*}
where again~$u=(1,\vec{n})$.
In the special case~$v=(1,\vec{0})$, where the gravitational system is at rest in the
reference frame distinguished by the regularization, this formula simplifies to
\begin{align}
\frac{\kappa(x)}{\kappa_0} = \frac{\kappa_\vol(x)}{\kappa_0} &= 1 + \frac{1}{2 \pi}
\int_{S^2} d\mu_{S^2}(\vec{n}) \int_{-\infty}^0
\big( u^j \partial_j \phi - 2 \partial_t \phi \big)(x+\alpha u) \: d\alpha \notag \\
&= 1 + 2 \phi(x) - 4 \int_{S^2} \frac{d\mu_{S^2}(\vec{n})}{4 \pi} \int_{-\infty}^0
\partial_t \phi(x+\alpha u) \:d\alpha \:, \label{newtonrest}
\end{align}
where in the last line we integrated by parts.
The second summand~$2 \phi(x)$ is a local contribution. It {\em{decreases}} 
the effective gravitational constant.
The last summand, however, {\em{increases}} the effective gravitational constant
in a collapsing scenario where~$\partial_t \phi$ and~$\phi$ have the same signs.

\subsection{Effects in the Solar System} \label{secsolar}
We now briefly discuss the effects in the solar system and estimate their order
of magnitude. We first consider the situation that the Sun is at rest in the
reference frame distinguished by the regularization.
Then the DGC function is described by~\eqref{newtonrest}. Moreover,
as the gravitational field of the Sun is static, the term~$\partial_t \phi$ vanishes, so that
\[ \frac{\kappa(x)}{\kappa_0} = \frac{\kappa_\vol(x)}{\kappa_0} = 1 + 2 \phi(x) \:. \]
This corresponds precisely to the {\em{red shift effect}} as described after~\eqref{locframe}.
Since the Schwarzschild radius of the Sun is~$r_\text{sch} \approx 2952 \,\text{m}$,
whereas the radius of the Sun is~$r_\text{sun} \approx 696 \,\text{km}$,
on the surface of the Sun we obtain the red shift
\beq \label{error}
2 \phi\big(r_\text{sun}) = \frac{2M}{r_\text{sun}} = \frac{r_\text{sch}}{r_\text{sun}}
\approx 4.23 \times 10^{-6}\:.
\eeq
The red shift effect due to the gravitational field of the Sun as observed on the Earth
is even much smaller. Namely, denoting the distance from the Earth to the Sun
by~$d_\text{es} \approx 149\times 10^6 \,\text{km}$, we obtain
\[ 2 \phi\big(d_\text{es} \big) \approx 1.97 \times 10^{-8}\:. \]
The red shift effect on the Earth generated by its own gravitational field is even smaller.
Namely, the Schwarzschild radius of the Earth is~$r_\text{sch} \approx 0.009 \,\text{m}$,
and since the radius of the Earth is~$r_\text{earth} \approx 6371 \,\text{km}$, we obtain
\[ \frac{r_\text{sch}}{r_\text{earth}} \approx 1.41 \times 10^{-9}\:. \]

For clarity, we point out that these deviations cannot be measured by an
E\"otv\"os-type experiment, because general relativity with DGC satisfies the
equivalence principle, so that inertial and gravitational mass coincide.
Instead, one must measure the value of the gravitational constant with high
accuracy using a {\em{Cavendish-type experiment}}.
Current experiments measure the gravitational constant
with a ``relative uncertainty of 150 parts per million''~\cite{Rosi:2014kva}.
Therefore, even if one could perform such an experiment on the surface of the Sun,
the effect~\eqref{error} would be by a factor~$1/35$ smaller than the accuracy of the experiment.
We conclude that measuring the red shift effect of DGC seems out of reach
of present-day technology.

Keeping in mind that the Sun moves relative to the center of our galaxy, and that our galaxy
moves relative to its galaxy cluster, etc., it seems physically reasonable that
in the reference frame distinguished by the regularization, the Sun is moving
with a constant velocity. In order to estimate the effect, let us consider the
limiting case~$\gamma=\infty$ where this velocity approaches the speed of light.
Then, according to~\eqref{gammainf}, the DGC function depends on the
position on the surface of the Sun. Apart from the red shift effect, also the
bending of null geodesics becomes important (see Figure~\ref{figangles}).
But again, the order of magnitude of the effects is governed by the quotient
in~\eqref{error}, being out of reach of present experiments.

\appendix
\section{The Geometry of Surfaces on the Light Cone} \label{appcone}
Let~$\scrS \subset \R^{1,3}$ be a smooth two-dimensional surface
lying on the future light cone of Minkowski space (i.e.~$\la x,x\ra_{\R^{1,3}}=0$ and~$x^0>0$
for all~$x \in \scrS$).
Assume that we can represent~$\scrS$ as the graph
over the unit sphere, i.e.\ that there is a diffeomorphism~$\Phi$ of the form
\[ \Lambda \::\: S^2 \subset \R^3
\rightarrow \scrS \subset \R^{1,3} \:,\qquad \Lambda(\vec{x}) = \lambda(\vec{x}) \begin{pmatrix} 1 \\ \vec{x} \end{pmatrix} \]
with a smooth function~$\lambda \in C^\infty(S^2, \R^+)$.
Next, we let~$\Phi : \Omega \subset \R^2 \rightarrow S^2$ be a parametrization of the unit sphere.
For example, choosing spherical coordinates~$(\vartheta, \varphi)$, the mapping~$\Phi$ is given by
\[ \Phi \::\: (0, \pi) \times (0, 2 \pi) \rightarrow S^2 \:,\qquad \Phi(\vartheta, \varphi) =
\begin{pmatrix} \sin \vartheta\; \cos \varphi \\ \sin \vartheta \; \sin \varphi \\ \cos \vartheta \end{pmatrix} \:, \]
but of course any other parametrization can be used just as well. We thus obtain a parametrization of~$\scrS$,
\[ \Xi := \Lambda \circ \Phi \::\: \Omega \subset \R^2 \rightarrow \scrS \subset \R^{1,3} \:, \qquad
\Xi(x) = \lambda \big( \Phi(x) \big) \begin{pmatrix} 1 \\ \Phi(x) \end{pmatrix} \:. \]
It follows that for any~$\alpha, \beta = 1,2$, the vectors
\beq \label{tangent}
\partial_\alpha \Xi(x) = \partial_\alpha \big( \lambda \circ \Phi \big)(x) \begin{pmatrix} 1 \\ \Phi(x) \end{pmatrix}
+ \big( \lambda \circ \Phi \big)(x) \begin{pmatrix} 0 \\ \partial_\alpha \Phi(x) \end{pmatrix}
\eeq
are tangential to~$\scrS$. Taking the Minkowski inner product of these tangent vectors,
one can make use of the fact that the first summand in~\eqref{tangent} is a null vector.
Moreover, the first summand is orthogonal to the second summand because
the vector~$\partial_\alpha \Phi(x)$ is tangential to the unit sphere, whereas the vector~$\Phi(x)$ is normal.
We thus obtain
\[  \big\la \partial_\alpha \Xi(x), \partial_\beta \Xi(x) \big\ra_{\R^{1,3}} = \big( \lambda \circ \Phi \big)(x)
\big\la \partial_\alpha \Phi(x),  \partial_\beta \Phi(x) \big\ra_{\R^3} \:. \]
Therefore, the induced metric on~$\scrS$ is conformal to the metric on the sphere,
\beq \label{induced}
ds^2 = \lambda(x)^2\: g_{S^2}(x) \:.
\eeq
Thus the volume measure of~$\scrS$ is given
\beq \label{mu2graph}
d\mu_\scrS = \lambda(x)^2\: d\mu_{S^2}(x) \:.
\eeq
It is worth noting that the induced metric~\eqref{induced} involves no derivatives of~$\lambda$.
The reason is that the terms involving first derivatives in~\eqref{tangent} drop out when taking the
Minkowski inner products.
The fact that~\eqref{mu2graph} involves no derivatives of~$\lambda$ makes it possible to
define the volume even in cases when the surface~$\scrS$ is not smooth. In fact, it suffices
to assume that~$\scrS$ is an $L^2$-graph over~$S^2$. Thus the volume is well-defined for
generalized two-dimensional surfaces of the form
\[ \scrS = \big\{ \lambda(x)\: x \;\big|\; x \in S^2 \subset \R^3 \text{ and } \lambda \in L^2(S^2, \R^+) \big\} \:. \]

We next compute the curvature of~$\scrS$. There are two
notions of curvature: intrinsic and extrinsic curvature. The intrinsic curvature is the Gaussian
curvature~$K$. It is computed by\footnote{We remark that the structure of this equation can be
understood directly from the Gau{\ss}-Bonnet theorem. Namely, combining~\eqref{scalsphere} with~\eqref{mu2graph},
we obtain
\[ \int_{\scrS} K\, d\mu_\scrS = \int_{S^2} \big( 1 - \Delta_{S^2} \log \lambda \big)\: d\mu_{S^2}
= 4 \pi - \int_{S^2} \Delta_{S^2} \log \lambda \: d\mu_{S^2} = 4 \pi \:, \]
where in the last step we applied the Gau{\ss} divergence theorem.}

\beq \label{scalsphere}
K =  \frac{1}{\lambda^2}\, \Big( 1 - \Delta_{S^2} \log \lambda \Big) \:,
\eeq
where~$\Delta_{S^2}$ is the Laplacian on~$S^2$ (see for example~\cite[page~237]{docarmo}).
In order to describe the extrinsic curvature, one must choose normal vector fields to~$\scrS$
in~$\R^{1,3}$. As~$\scrS$ is two-dimensional and Riemannian, we can choose one time-like
normal~$\nu_0$ and one space-like normal~$\nu_1$, for convenience pseudo-orthonormalized, i.e.
\[ \la \nu_0, \nu_0 \ra_{\R^{1,3}} = 1 \:,\quad \la \nu_0, \nu_1 \ra_{\R^{1,3}} = 0
\:,\quad \la \nu_1, \nu_1 \ra_{\R^{1,3}} = -1\:. \]
Then in a local parametrization~$\Xi$, the second fundamental form can be introduced as
a Minkowski vector by
\beq \label{Habdef}
H_{\alpha \beta}
= \nu_0 \:\la \nu_0, \partial_\alpha e_\beta \ra_{\R^{1,3}} - \nu_1\: \la \nu_1, \partial_\alpha e_\beta \ra_{\R^{1,3}}
\eeq
(where~$e_\alpha = \partial_\alpha \Xi$ are the tangent vectors).
The mean curvature vector~$H$ is defined by
\beq \label{Hdef}
H = g^{\alpha \beta}\: H_{\alpha \beta} \:.
\eeq
The next lemma shows the only scalar quantity which can be formed of the second fundamental form
is the Gaussian curvature.
\begin{Lemma}  \label{lemmaGauss}
The square of the second fundamental form
and the Minkowski length of the mean curvature vector are related to the Gaussian curvature by
\beq \label{HabHab}
\la H, H \ra_{\R^{1,3}} = -4K \qquad \text{and} \qquad
\la H_{\alpha \beta}, H^{\alpha \beta} \ra_{\R^{1,3}} = -2 K \:.
\eeq
\end{Lemma}
\Proof 
It is most convenient to work in the local parametrizations
$\Phi : (-1,1)^2 \rightarrow S^2$ and~$\Xi : (-1,1)^2 \rightarrow \scrS$ given by
\[ \Phi(x,y) = \lambda(x,y) \, \begin{pmatrix} x \\ y \\[0.2em] \sqrt{1-x^2-y^2} \end{pmatrix} \:,\qquad
\Xi(x,y) = \lambda(x,y) \begin{pmatrix} 1 \\ \Phi(x,y) \end{pmatrix} \:. \]
Clearly, it suffices to derive the identities~\eqref{HabHab} at the point~$p = \Xi(0,0)$.
A short calculation shows that
\[ g_{\alpha \beta}(p) = \lambda(0,0)^2\: \delta_{\alpha \beta} \:. \]
We next introduce the Minkowski vector
\begin{align*}
L &= g^{\alpha \beta}(p) \:\partial_{\alpha \beta} \Xi(0,0) = \frac{1}{\lambda(0,0)^2}
\big( \partial_{xx} + \partial_{yy} \big) \Xi(0,0) \\
&= \frac{1}{\lambda^2(0,0)} \begin{pmatrix} \partial_{xx} \lambda + \partial_{yy} \lambda \\
2 \partial_x \lambda \\
2 \partial_y \lambda \\
-2 \lambda + \partial_{xx} \lambda + \partial_{yy} \lambda \end{pmatrix} \bigg|_{(0,0)}\:.
\end{align*}
By direct computation, one verifies that~$L$ is orthogonal to~$T_p\scrS$, i.e.
\[ \la L, \partial_x \Phi \ra_{\R^{1,3}} = \la L, \partial_y \Phi \ra_{\R^{1,3}} = 0 \:. \]
Hence~$T$ is a linear combination of the normals~$\nu_0$ and~$\nu_1$. 
Using this fact in~\eqref{Habdef} and~\eqref{Hdef}, one finds that~$H=L$.
Computing the Minkowski inner product gives
\beq \label{HH}
\la H,H \ra = -\frac{4}{\lambda(0,0)^2} \:\Big( 1 - \partial_{xx} \log \lambda - \partial_{yy} \log \lambda \Big)
\Big|_{(0,0)}\:.
\eeq
Next, a short computation shows that the parametrization~$\Phi$
gives a Gaussian coordinate system around~$\Phi(0,0)$. Therefore, the
second derivatives in~\eqref{HH} can be rewritten as~$\Delta_{S^2} \log \lambda$.
Comparing with~\eqref{scalsphere}, we obtain the first equation in~\eqref{HabHab}.

The second equation in~\eqref{HabHab} can be derived similarly by a straightforward computation.
\QED

\section{The Regularized Hadamard Expansion} \label{apphadamard}
We consider the setting in~\cite{lqg} and also use the same notation.
Thus let~$(\scrM, g)$ be a globally hyperbolic space-time.
We choose a global time function~$\mathfrak{t}$.
Let~$\Omega \subset \scrM$ be a geodesically
convex subset (see~\cite[Definition~1.3.2]{baer+ginoux}). Then for any~$x, y \in \Omega$, there is a
unique (unparametrized) geodesic~$\gamma$ in~$\Omega$ joining~$y$ and~$x$.
We denote the spinorial Levi-Civita parallel transport along~$\gamma$ by
\[ \Pi^y_x : S_y\scrM \rightarrow S_x\scrM \:. \]
Moreover, we denote the squared length of this geodesic by
\[ \Gamma(x,y) = \pm \left( \int \sqrt{\big| \la \dot{\gamma}(\tau), \dot{\gamma}(\tau) \ra_{\gamma(\tau)}
\big| }\: d\tau \right)^2 \]
(where~$\tau$ is any parametrization), where we choose the plus sign in timelike directions
and the minus sign in spacelike directions.
In the next lemma we collect a few elementary properties of~$\Pi$ and~$\Gamma$.
\begin{Lemma} \label{lemmaGamma}
Let~$\gamma$ be the geodesic joining~$x$ and~$y$, parametrized such that~$\gamma(0)=y$
and~$\gamma(1)=x$. Then
\beq \label{gradGamma}
\grad_x \Gamma(x,y) = 2\, \dot{\gamma}(1) \qquad \text{and} \qquad
\grad_y \Gamma(x,y) = -2\, \dot{\gamma}(0) \:.
\eeq
Moreover, the following identities hold:
\begin{align}
\big\la \grad_x\Gamma,\grad_x\Gamma(x,y) \big\ra_x &= 4 \Gamma(x,y) \label{gradG} \\
g^{jk}\: \frac{\partial \Gamma(x,y)}{\partial x^j}\: \nabla_{x^k} \Pi^y_x &= 0 \:. \label{Piparallel}
\end{align}
\end{Lemma}
\Proof  We denote the velocity vector of~$\gamma$ at~$y$ by~$u := \dot{\gamma}(0) \in T_y \scrM$.
The geodesic equation implies that the inner
product~$\la \dot{\gamma}(\tau), \dot{\gamma}(\tau) \ra_{\gamma(\tau)}$ is a constant.
This makes it possible to write~\eqref{Gammadef} as
\beq \label{Gammarel}
\Gamma(x,y) = \la u, u \ra_y \:.
\eeq
Next, we let~$\exp_y : T_y\scrM \rightarrow \scrM$ be the exponential map.
Then clearly~$\exp_y u = x$. Moreover, \eqref{Gammarel} can be written as
\beq \label{Gammarel2}
\Gamma(x,y) = \big\la \exp_y^{-1}(x), \exp_y^{-1}(x) \big\ra_y \:.
\eeq
Since~$\Omega$ is totally geodesic, the inverse of this mapping exists on~$\Omega$,
\[ \exp_y^{-1} \::\: \Omega \rightarrow T_y \scrM \:. \]
Differentiating the identity~$\exp_y^{-1} \circ \exp_y = \1_{T_y\scrM}$ at~$u$, we obtain
\[ D\exp_y^{-1}\big|_x = \left( D\exp_y \big|_u \right)^{-1} \:. \]
Hence, differentiating~\eqref{Gammarel2} we obtain for any~$v \in T_x\scrM$
\[ D_x \Gamma(x,y)\cdot v = 2 \:\big\la u, D \exp_y^{-1}\big|_x v \big\ra_y = 
2 \:\big\la u,  \left( D\exp_y \big|_u \right)^{-1} v \big\ra_y \:. \]
In order to simplify this relation, we make use of the Gauss lemma, stating that
(see for example~\cite[Lemma~3.5]{docarmo})
\[ \la \dot{\gamma}(1), D\exp_y \big|_u w \ra_x = \la u, w \ra_y \qquad
\text{for all~$w \in T_y\scrM$}\:. \]
Applying this lemma for~$w = ( D\exp_y \big|_u)^{-1} v$, we obtain
\[ D_x \Gamma(x,y)\cdot v = 2 \:\big\la \dot{\gamma}(1),  v \big\ra_x \:, \]
proving the first equation in~\eqref{gradGamma}. The second follows immediately
by interchanging the roles of~$x$ and~$y$ and reparametrizing~$\gamma$.

The first equation in~\eqref{gradGamma} implies that
\[ \big\la \grad_x\Gamma(x,y),\grad_x\Gamma(x,y) \big\ra_x = 4 \: \big\la \dot{\gamma}(1), \dot{\gamma}(1) \big\ra_x
= 4 \: \big\la \dot{\gamma}(0), \dot{\gamma}(0) \big\ra_y = 4 \: \la u,u \ra_y\:, \]
and applying~\eqref{Gammarel} gives~\eqref{gradG}.

The relation~\eqref{Piparallel} follows by substituting~\eqref{gradGamma} and using the fact that
the tangential covariant derivative of the parallel transport along the curve~$\gamma$ is zero.
\QED
Moreover, parametrizing the geodesic~$\gamma$ such that~$\gamma(0)=y$
and~$\gamma(\tau)=x$, we have
\beq \label{gradG2}
\grad_x \Gamma(x,y) = 2 \tau\: \frac{d}{d\tau} \gamma(\tau) \:.
\eeq
In order to prescribe the behavior of the singularities on the light cone, we set
\[ \Gamma_\varepsilon(x,y) = \Gamma(x,y) + i \varepsilon \big(\mathfrak{t}(x) - \mathfrak{t}(y) \big) \]
and introduce the short notation
\[ \frac{1}{\Gamma^p} = \lim_{\varepsilon \searrow 0} \frac{1}{(\Gamma_\varepsilon)^p} \qquad \text{and} \qquad
\log \Gamma = \lim_{\varepsilon \searrow 0} \log \Gamma_\varepsilon 
= \log |\Gamma| - i \pi\, \epsilon\big(\mathfrak{t}(x) - \mathfrak{t}(y) \big) \]
(where~$\epsilon$ denotes the step function~$\epsilon(x)=1$ if~$x>1$ and~$\epsilon(x)=-1$ otherwise),
with convergence in the distributional sense.
Here the logarithm is cut along the positive real axis, with the convention
\[ \lim_{\varepsilon \searrow 0} \log(1+i \varepsilon) = - i \pi\:. \]
Moreover, we let~$\vleck(x,y)$ be the square root of the van Vleck-Morette determinant
(see for example~\cite{moretti}). In normal coordinates around $y$, it is given by 
\[ \vleck(x,y)=|\det(g(x))|^{-\frac{1}{4}} \:. \]
Then the unregularized fermionic projector has
a singularity on the light cone of the form (see~\cite[Corollary~5.7]{lqg})
\begin{align*}
P(x,y) &= \frac{1}{8 \pi^3}\;
\frac{i \vleck(x,y)}{\Gamma(x,y)^2} \:\big( \grad_x\Gamma(x,y) \big) \cliff \Pi^y_x + \O \Big(\frac{1}{\Gamma} \Big) \\
&= -\frac{1}{8 \pi^3}\; \Dir_x \bigg( \frac{\vleck(x,y)}{\Gamma(x,y)} \:\Pi^y_x \bigg) + \O \Big(\frac{1}{\Gamma} \Big) \:.
\end{align*}
In order to describe the lower orders on the light cone systematically, it is useful to make the ansatz
(see~\cite{sahlmann2001microlocal} or~\cite[page~156]{hack})
\[ P(x,y) = \Dir_x \left( \frac{U(x,y)}{\Gamma(x,y)}
+ V(x,y)\: \log \Gamma(x,y) + W(x,y) \right) , \]
where~$U$, $V$ and~$W$ are smooth functions on~$\scrM \times \scrM$
taking values in the~$4 \times 4$-matrices acting on the spinors. Expanding these functions about the light cone
\[ U(x,y) = \sum_{n=0}^\infty U_n(x,y)\: \Gamma(x,y)^n \]
(and similarly for~$V(x,y)$ and~$W(x,y)$, one sees that
\[ U_0(x,y) = -\frac{1}{8 \pi^3}\: \vleck(x,y)\: \Pi^y_x \:. \]
For the regularized fermionic projector we set
\[ \Gamma_\varepsilon(x,y) =
\Gamma(x,y) + i \varepsilon f(x,y) \]
with a smooth function~$f(x,y)$. We make the ansatz
\[ P^\varepsilon(x,y)
= \Dir_x \left( \frac{U(x,y)}{\Gamma_\varepsilon(x,y)}
+ V(x,y)\: \log \Gamma_\varepsilon(x,y) + W(x,y) \right)\bigg( 1 + \O\Big( \frac{\varepsilon^2}
{\Gamma_\varepsilon(x,y)} \Big) \bigg) \:. \]
Thus the leading singularity is of the form
\[ P^\varepsilon(x,y) = -\frac{1}{8 \pi^3}\: \Dir_x \left( \frac{\vleck}{\Gamma_\varepsilon}
\: \Pi^y_x \right) \bigg( 1 + \O\Big( \frac{\varepsilon^2}{\Gamma_\varepsilon} \Big) \bigg)
+ \O\big( \Gamma_{\varepsilon}^{-1} \big) \:. \]
Applying the Dirac operator to the leading term gives
\begin{align*}
-\Dir^2_x \left( \frac{\vleck}{\Gamma_\varepsilon}
\: \Pi^y_x \right) &= \Delta^{\text{\tiny{$S\scrM$}}}_x \left( \frac{\vleck}{\Gamma_\varepsilon} \: \Pi^y_x \right) + \O\big( \Gamma_{\varepsilon}^{-1} \big)\\
&= \Delta_x \left( \frac{\vleck}{\Gamma_\varepsilon} \right) \:\Pi^y_x
+ 2  g^{jk}\: \frac{\partial}{\partial x^j} \left( \frac{\vleck}{\Gamma_\varepsilon} \right) \;\nabla_{x^k} \Pi^y_x + \O\big( \Gamma_{\varepsilon}^{-1} \big) \:,
\end{align*}
where we applied the Bochner-Lichnerowicz-Weitzenb\"ock formula
(and~$\Delta^{\text{\tiny{$S\scrM$}}}$ is the Laplacian on the spinor bundle). Next,
using the convention that all derivatives are carried out with respect to the variable~$x$,
\begin{align}
\partial_j \left( \frac{\vleck}{\Gamma_\varepsilon} \right)
&= \frac{\partial_j \vleck}{\Gamma_\varepsilon} - \frac{\vleck}{\Gamma_\varepsilon^2}
\:\Big( \partial_j \Gamma + i \varepsilon\, \partial_j f \Big) \notag \\
\Delta \left( \frac{\vleck}{\Gamma_\varepsilon} \right)
&= \frac{\Delta \vleck}{\Gamma_\varepsilon} + \frac{2 \vleck}{\Gamma_\varepsilon^3}
\:\Big( \partial_j \Gamma + i \varepsilon \,\partial_j f \Big)\Big( \partial^j \Gamma + i \varepsilon \,\partial^j f \Big) \notag \\
&\qquad - \frac{\vleck}{\Gamma_\varepsilon^2} \:\Big( \Delta \Gamma + i \varepsilon \Delta f \Big)
- \frac{\partial^j \vleck}{\Gamma_\varepsilon^2} \:\Big( \partial_j \Gamma + i \varepsilon \partial_j f \Big) \notag \\
&\!\!\!\overset{\eqref{gradG}}{=} \frac{\Delta \vleck}{\Gamma_\varepsilon} + \frac{2 \vleck}{\Gamma_\varepsilon^3}
\:\Big( 4 \Gamma + 2 i \varepsilon \,(\partial^j f)
(\partial_j \Gamma) - \varepsilon^2\, (\partial_j f) (\partial^j f) \Big) \notag \\
&\qquad - \frac{\vleck}{\Gamma_\varepsilon^2} \:\Big( \Delta \Gamma + i \varepsilon \,\Delta f \Big)
- \frac{\partial^j \vleck}{\Gamma_\varepsilon^2} \:\Big( \partial_j \Gamma + i \varepsilon \,\partial_j f \Big) \notag \\
&= \frac{4 \vleck}{\Gamma_\varepsilon^3} \: i \varepsilon\, \Big( -2 f + (\partial^j \Gamma) (\partial_j f) \Big) \label{im1} \\ 
&\qquad - \frac{1}{\Gamma_\varepsilon^2} \:i \varepsilon\, \Big( \vleck\, \Delta f
+ (\partial^j \vleck) (\partial_j f) \Big) \label{im2} \\
&\qquad + \frac{1}{\Gamma_\varepsilon^2} \Big( 8 \vleck
- (\partial^j \Gamma) (\partial_j \vleck)  - \vleck\:\Delta \Gamma \Big) \label{re1} \\
&\qquad +\frac{\Delta \vleck}{\Gamma_\varepsilon} 
- \frac{2 \vleck}{\Gamma_\varepsilon^2}\: \frac{\varepsilon^2}{\Gamma_\varepsilon}\,
(\partial_j f) (\partial^j f) \label{re2} \\
2  g^{jk}\: \frac{\partial}{\partial x^j} &\left( \frac{\vleck}{\Gamma_\varepsilon} \right) \;\nabla_{x^k} \Pi^y_x
= - \frac{2}{\Gamma_\varepsilon^2}\: \partial^j \big( \Gamma(x,y) + i \varepsilon f(x,y) \big) \;\nabla_j \Pi^y_x \notag \\
&=-\frac{2}{\Gamma_\varepsilon^2}\: \bigg( (\partial^j \Gamma)\: \nabla_j \Pi^y_x
+ i \varepsilon \,(\partial^j f)\: \nabla_j \Pi^y_x \bigg)
\overset{\eqref{Piparallel}}{=} -\frac{2  i \varepsilon}{\Gamma_\varepsilon^2}\:\big(\partial^j f \big)\;\nabla_j \Pi^y_x
\label{im3} \:.
\end{align}
Here we wrote the real and imaginary parts separately. For the imaginary part, the term~\eqref{im1}
is the leading singularity, whereas the contributions~\eqref{im2} and~\eqref{im3} are of higher order
in~$\Gamma_\varepsilon$. For the real part, the leading singularity is given by~\eqref{re1},
whereas the terms in~\eqref{re2} are error terms. Namely, the first summand in~\eqref{re2}
is of higher order in~$\Gamma_\varepsilon$, and the second summand is of higher order
in~$\varepsilon^2/\Gamma_\varepsilon$. We thus obtain the equations for~$\vleck$ and~$f$
\begin{align}
(\partial_j \Gamma) (\partial^j \vleck) &=  8 \vleck - \vleck\:\Delta \Gamma \label{transport1} \\
(\partial^j \Gamma) (\partial_j f)  &= 2 f \:. \label{transport2}
\end{align}
The equation~\eqref{transport1} is the usual transport equation for the 
van Vleck-Morette determinant. The novel feature of the regularized Hadamard expansion is the
transport equation for~$f$. In order to solve this transport equation, we consider a geodesic~$\gamma$
with~$\gamma(0)=y$ and use~\eqref{gradG2} to obtain
\[  \tau \:\dot{\gamma}^j(\tau)  \:\partial_j f \big( \gamma(\tau) \big) = f \big( \gamma(\tau) \big) \]
and thus
\[ \tau\: \frac{d}{d\tau} f \big( \gamma(\tau) \big) = f \big( \gamma(\tau) \big) \:. \]
Setting~$h(\tau) =  h(\gamma(\tau))$, we obtain the ordinary differential equation
\[ \tau\: \frac{d}{d\tau} h(\tau) = h(\tau) \:, \]
having the general solution~$h(\tau) = c \tau$ with a free real parameter~$c$.
We conclude that~$f$ has the general form
\[ f \big( \gamma(\tau) \big) = c \tau \:. \]
Hence along a geodesic through~$y$, the function~$f(.,y)$ simply is an affine parameter
along this geodesic with~$f(y,y)=0$.

\section{The DGC Tensor} \label{appE}
Taking the divergence of the Einstein equations with DGC~\eqref{EDGC} but without the DGC tensor~$E_{jk}$ gives
\begin{align*}
\nabla^j \Big( R_{jk} - \frac{1}{2}\:R \: g_{jk} + \Lambda\, g_{jk} -  \kappa(p)\: T_{jk} \Big)
= -\big( \partial^j \kappa \big) \: T_{jk} \:.
\end{align*}
Therefore, our task is to construct a tensor field~$E_{jk}$ such that
\beq \label{divE}
\nabla^j E_{jk} = -\big(\partial^j \kappa \big) \: T_{jk} \:.
\eeq
In preparation, we introduce some notation. We let~$\ell_{jk}$ be the symmetric tensor
\[ \ell_{jk}(p) = \int_{D_p\scrL} x_j\, x_k\: d\mu_{D_p\scrL}(x)\:. \]
This tensor is trace-free, because
\beq \label{notrace}
g^{jk}(p)\: \ell_{jk}(p) = \int_{D_p\scrL} \la x, x \ra_p \: d\mu_{D_p\scrL}(x) = 0 \:.
\eeq
In the example~\eqref{LMink} in Minkowski space, one obtains~$(\ell_{jk}) = 8 \pi^2/(3 \varepsilon^2)
\: \text{diag}(3,1,1,1)$. Therefore, it seems reasonable to assume that~$\ell_{jk}$ is invertible,
meaning that there is a tensor~$\ell^{-1}_{jk}$ with
\[  \ell^{ij}\: \ell^{-1}_{jk} = \delta^i_k\:. \]

The DGC tensor~$E_{jk}$ will involve integrals over null geodesics.
Using the exponential map, the parametrized geodesic~$\gamma(\tau)$ with~$\gamma(0)=p$
and~$\dot{\gamma}(0)=x$ can be written conveniently as
\beq \label{gexp}
\gamma(\tau) = \exp_p(\tau x)\:.
\eeq
The integrals over the null geodesics can go either to the past or the future.
Since we are mainly interested in the Cauchy problem, we assume that~$\scrM$ is
time-orientable and integrate along the null geodesics to the past
(integrating instead to the future is a special case of the freedom in modifying
the DGC tensor as discussed in Remark~\ref{remfreedom} below).
We denote all future-directed vectors in~$D_p\scrL$ by~$D_p\scrL^\vee$.

We now introduce the {\em{DGC tensor}}~$E_{jk}$ by
\beq \label{Edef}
\begin{split}
E_{jk}(p) &= -\int_{D_p\scrL^\vee} d\mu_{D_p\scrL}(x)\: x_j\, x_k\, x_l \\
&\qquad \times
\int_{\tau_{\min}}^0 d\tau \:\big(\Pi_{p, \exp_p(\tau x)} \big)^{la}
\: \Big( \ell^{-1}_{ab} \:\big(\partial_c \kappa \big) \: T^{bc} \Big) \Big|_{\exp_p(\tau x)}\:,
\end{split}
\eeq
where~$\Pi_{p, \exp_p(\tau x)}$ is the Levi-Civita parallel transport along the geodesic from~$\exp_p(\tau x)$
to~$p$, and~$\tau_{\min} \in [-\infty, 0)$ is the minimal~$\tau$ for which the exponential map is defined
(thus~$\tau_{\min}=-\infty$ if the null geodesic exists for all negative values of the affine parameter,
but it is finite for example if the geodesic hits the big bang singularity).

\begin{Prp} The DGC tensor~\eqref{Edef} is trace-free and satisfies the divergence equation~\eqref{divE}.
\end{Prp}
\Proof Exactly as explained in~\eqref{notrace}, one sees immediately that the DGC tensor
is trace-free. In order to compute its divergence, it is most convenient to choose a Gaussian
reference frame at~$p$. Representing~$D_p\scrL$ as in Appendix~\ref{appcone} by a graph,
we can write~\eqref{Edef} as (cf.~\eqref{mu2graph})
\begin{align*}
E_{jk}(p) &= -\int_{S^2} \lambda^2\: d\mu_{S^2}(\vec{n}) \; \lambda^3 \, u_j\, u_k\, u_l \\
&\qquad \times \int_{\tau_{\min}}^0 d\tau \:\big(\Pi_{p, \exp_p(\tau x)} \big)^{la}\,
\Big( \: \ell^{-1}_{ab} \:\big(\partial_c \kappa \big) \: T^{bc} \Big) \Big|_{\exp_p(\tau \lambda u)}\:,
\end{align*}
where we used~\eqref{mu2graph} and set~$u(\vec{n}) = (1, -\vec{n})/\varepsilon$.
We now take the covariant divergence~$\nabla^j E_{jk}(p)$.
Clearly, the factors~$\lambda$ depend on~$\vec{n}$ and~$p$. Differentiating them with respect to~$p$
gives combinations~$u^j \partial_\lambda$. However, as the vector~$x=\lambda u$ is parallel along
the geodesic and the derivative of~$u$ at~$p$ vanishes in our Gaussian coordinate system,
we conclude that~$u^j \partial_j \lambda=0$. Thus it remains to differentiate the integral.
Since the resulting derivative is tangential to the light cone, it can be rewritten
as a derivative with respect to~$\tau$. We thus obtain
\begin{align*}
\lambda u^j & \frac{\partial}{\partial p^j} \int_{\tau_{\min}}^0 
\big(\Pi_{p, \exp_p(\tau x)} \big)^{la} \,\Big( \: \ell^{-1}_{ab} \:\big(\partial_c \kappa \big) \: T^{bc} \Big) \Big|_{\exp_p(\tau \lambda u)} \:d\tau \\
&= \frac{d}{ds} \int_{\tau_{\min}-s}^0
\big(\Pi_{p, \gamma(\tau+s)} \big)^{la}\,
\Big( \ell^{-1}_{ab} \:\big(\partial_c \kappa \big) \: T^{bc} \Big) \Big|_{\gamma(\tau+s)} \:d\tau \bigg|_{s=0} \\
&= \frac{d}{ds} \int_{\tau_{\min}}^s
\big(\Pi_{p, \gamma(\tau)} \big)^{la}\,
\Big( \ell^{-1}_{ab} \:\big(\partial_c \kappa \big) \: T^{bc} \Big) \Big|_{\gamma(\tau)} \:d\tau \bigg|_{s=0} \\
&= g^{la}(p)\: \ell^{-1}_{ab}(p) \:\big(\partial_c \kappa(p) \big) \: T^{bc}(p)\:.
\end{align*}
It follows that
\begin{align*}
\nabla^j E_{jk}(p) &= -\int_{S^2} \lambda^2\: d\mu_{S^2}(\vec{n}) \; \lambda^2 \, u_k\, u^a
\:\ell^{-1}_{ab}(p) \:\big(\partial_c \kappa(p) \big) \: T^{bc}(p) \\
&= -\ell^{-1}_{ab}(p) \:\big(\partial_c \kappa(p) \big) \: T^{bc}(p)
\int_{D_p\scrL^\vee} x_k\, x^a\: d\mu_{D_p\scrL}(x) \\
&= -\ell^{-1}_{ab}(p) \:\big(\partial_c \kappa(p) \big) \: T^{bc}(p)\; \ell_k^a
= -\big(\partial_c \kappa(p) \big) \: T_k^{\;\,c}(p) \:,
\end{align*}
concluding the proof.
\QED

\begin{Remark} \label{remfreedom} {\em{
We finally comment on the freedom in choosing the DGC tensor.
The formula for the DGC tensor given in~\eqref{Edef} follows naturally
by solving the divergence equation~\eqref{divE} for an expression formed of
integrals along geodesics and integrals over~$D_p\scrL$.
The expression in~\eqref{Edef} does not involve free parameters, and the authors do not see
how this expression could be modified.
The only obvious freedom is that every line integral in~\eqref{Edef} comes with an integration constant.
In~\eqref{Edef}, this integration constant was determined by demanding that the line integral should vanish
at~$\tau_{\min}$.
But clearly, this was an arbitrary choice, motivated by the wish to have simple initial data in the early universe.
The freedom in choosing the integration constants can be described systematically by
adding to~$E_{jk}$ the tensor
\[ \int_{D_p\scrL^\vee} d\mu_{D_p\scrL}(x)\: x_j\, x_k\, f(p,x) \]
with a function~$f : D\scrL \rightarrow \R$ which is constant along null geodesics, i.e.
\[ x^j \frac{\partial}{\partial p^j} f(p,x) = 0 \:. \]
For example, one could choose~$f$ such that the DGC tensor vanishes in the present universe,
but then it would be non-zero at the big bang singularity.
}} \QEDrem
\end{Remark}

\Thanks {{\em{Acknowledgments:}}
We would like to thank H{\r{a}}kan Andr{\'e}asson, Niky Kamran,
Olaf M\"uller and Gerhard Rein for helpful discussions.
F.F.\ is grateful to the Center of Mathematical Sciences and Applications at
Harvard University for hospitality and support. 


\providecommand{\bysame}{\leavevmode\hbox to3em{\hrulefill}\thinspace}
\providecommand{\MR}{\relax\ifhmode\unskip\space\fi MR }
\providecommand{\MRhref}[2]{%
  \href{http://www.ams.org/mathscinet-getitem?mr=#1}{#2}
}
\providecommand{\href}[2]{#2}


\end{document}